\documentclass[twocolumn,english]{revtex4-1}
\usepackage[T1]{fontenc}
\usepackage[latin9]{inputenc}
\usepackage{verbatim}
\usepackage{calc}
\usepackage{amsmath}
\usepackage{amssymb}
\usepackage{cancel}
\usepackage{graphicx}

\makeatletter
\usepackage{pdfsync}

\makeatother

\usepackage{babel}
\begin{document}
\global\long\def\ket#1{\left|#1\right\rangle }%

\global\long\def\bra#1{\left\langle #1\right|}%

\global\long\def\braket#1#2{\left\langle #1\left|#2\right.\right\rangle }%

\global\long\def\ketbra#1#2{\left|#1\right\rangle \left\langle #2\right|}%

\global\long\def\braOket#1#2#3{\left\langle #1\left|#2\right|#3\right\rangle }%

\global\long\def\mc#1{\mathcal{#1}}%

\global\long\def\nrm#1{\left\Vert #1\right\Vert }%

\title{The Passivity Deformation Approach for the Thermodynamics of Isolated
Quantum Setups}
\author{Raam Uzdin}
\affiliation{Fritz Haber Research Center for Molecular Dynamics, Hebrew University
of Jerusalem, Jerusalem 9190401, Israel}
\email{raam@mail.huji.ac.il}

\author{Saar Rahav}
\affiliation{Schulich Faculty of Chemistry, Technion--Israel Institute of Technology,
Haifa 3200008, Israel}
\email{rahavs@technion.ac.il}

\selectlanguage{english}%
\begin{abstract}
Recently implemented quantum devices such as quantum processors and
quantum simulators combine highly complicated quantum dynamics with
high-resolution measurements. We present a passivity deformation methodology
that sets thermodynamic constraints on the evolution of such quantum
devices. This framework enhances the thermodynamic predictive power
by simultaneously resolving four of the cardinal deficiencies of the
second law in microscopic setups: i) It yields tight bounds even when
the environment is microscopic; ii) The ultra-cold catastrophe is
resolved; iii) It enables to integrate conservation laws into thermodynamic
inequalities for making them tighter; iv) it bounds observables that
are not energy-based, and therefore do not appear in the second law
of thermodynamics. Furthermore, this framework provides insights to
non-thermal environments, correlated environments, and to coarse-graining
in microscopic setups. Our findings can be explored and used in physical
setups such as trapped ions, superconducting circuits, neutral atoms
in optical lattices and more.
\end{abstract}
\maketitle

\section{Introduction}

The theory of thermodynamics emerged during the industrial revolution.
This celebrated theory was developed due to the pressing need to know
how much coal a steam engine requires to accomplish a task. One of
the strengths of the theory is its ability to make predictions that
do not depend on the precise details of a specific engine. Instead,
it provides universal laws (bounds) that apply to all systems and
processes.

Rapid technological advances allow us unprecedented ability to control
and manipulate setups with highly pronounced quantum dynamics. Examples
include dozens of interacting (atomic) spins in ion traps, neutral
atoms in optical lattices, superconducting circuits, Rydberg atom
lattices, etc. These setups are candidates for the realization of
quantum technologies such as computation, communication, and more.
Such applications require the ability to measure very specific observables
that would, for instance, correspond to the result of a quantum computation.
For example, in ion traps or superconducting circuits, it is possible
to measure quantities such as the polarization of specific spins,
their mutual polarization covariance (correlation), or the population
of some preferred states. We refer to such observables as ``fine-grained''
to distinguish them from observables that characterize the whole system,
such as energy, volume, entropy, etc.

Since fine-grained experimental data is presently available, it is
desirable to have a theory that can make predictions about such quantities.
One possible approach is to model all the details of a setup and solve
or simulate the process of interest. Such an approach has to be repeated
if the setup is driven using a different protocol or a different initial
condition. Furthermore, such an approach is typically unfeasible in
quantum computations and simulators that attempt to solve problems
that are computationally hard (classically). A different approach,
more in the spirit of thermodynamics, is to identify constraints that
are applicable to a whole class of processes without the need to explicitly
solve for the evolution. This is the approach we take in the present
paper.

\begin{figure}
\includegraphics[width=8.6cm]{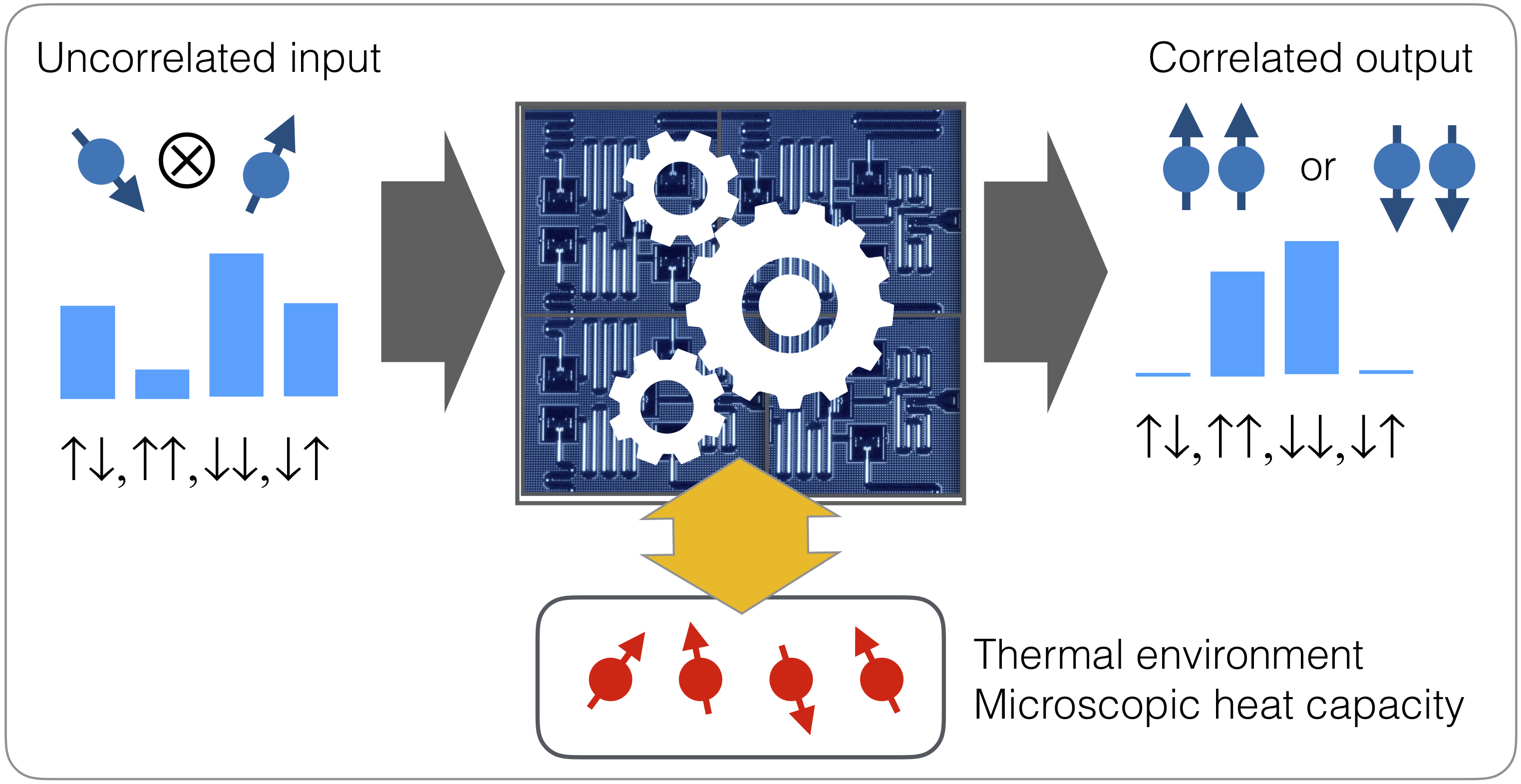}

\caption{\label{fig: X machine}An example of an exotic heat machine. A small,
initially thermal environment composed of four spins is used to make
a two-spin system more correlated (either $00$ or $11$). Although
the second law is applicable it is not the correct tool to provide
a performance limit on this fine-grained task.}
\end{figure}

The utility of thermodynamic-like bounds on fine-grained quantities
can also be illustrated using the example depicted in Fig. \ref{fig: X machine}.
The figure depicts a setup of six spins that are initially prepared
in a thermal state characterized by an inverse temperature $\beta$.
One then drives the system unitarily with the goal of increasing the
probability that spins 1 and 2 are aligned. The probability of this,
$p_{00}+p_{11}$, is associated with the expectation value of the
operator $A=\ketbra{0_{1}0_{2}}{0_{1}0_{2}}+\ketbra{1_{1}1_{2}}{1_{1}1_{2}}$
(The identify over spins 3-6 is implied). A more quantum fine-grained
task would be to increase the expectation value of some entanglement
witness. Since $\left\langle A\right\rangle $ is not the energy of
a subsystem the second law of thermodynamics can not be used to obtain
a useful bound on the changes of $\left\langle A\right\rangle $.
In particular, it is of interest to understand how much heat and work
are needed for changing $\left\langle A\right\rangle $ and how does
the performance depend on the size of the small environment and its
initial temperature. The approach presented in this paper, allows
the derivation of such bounds.

Our approach, \emph{passivity deformation}, uses the recently introduced
notion of global passivity \citep{GlobalPassivity} as a starting
point, yet it quickly deviates from global passivity in order to overcome
its inherent limitations. Global passivity produces a family of inequality
on observables, one of which is the second law. For the reader familiar
with thermodynamic resource theory \citep{Lostaglio2019RTreview,BrandaoPnasRT2ndLaw,horodecki2013fundamental,GourRTreview,LostaglioRudolphCohConstraint},
we point out that the global passivity inequalities are different
because they deal only with observable quantities. Very recently the
predictive power of global passivity bounds was experimentally demonstrated
using the IBM superconducting quantum processors \citep{uzdin2019IBMexp}.
By checking the validity of these inequalities it was possible to
detect heat leaks that the second law and other thermodynamic frameworks
could not detect. In \citep{GlobalPassivity} they were also used
to detect subtle Maxwell demons (i.e. weak feedback operations) in
numerical simulations.

Unfortunately, global passivity is not free of flaws. Despite being
based on a different mathematical framework, global passivity still
posses the four major deficiencies of the second law in microscopic
setups. The first, as discussed above, is the inability to address
fine-grained quantities. The second deficiency is the ``ultra cold''
catastrophe'', in which the second law becomes trivial and useless
when one of the environments is very cold. Specifically, as $T_{c}\to0$
the $Q_{c}/T_{c}$ term in the Clausius inequality form of the second
law dominates the inequality and the second law reduces to $Q_{c}\ge0$.
That is, the energy of the cold bath cannot be decreased. However,
when $T_{c}$ is low enough the environment starts in its ground state
so clearly the average energy cannot go further down. Thus, the second
law provides no useful information in this case. See \citet{Paternostro2017EntProdWigner,Landi2019landauer_zero_temp,Landi_Goes2019zero_temp}
for additional interesting approaches for handling the ultra-cold
catastrophe. The third deficiency is that the second law and the global
passivity bounds are unattainable when the environments go out of
equilibrium. In microscopic setups where the heat capacities are very
small this a major drawback that degrades the predictive power of
these bounds. Finally, the forth deficiency is the inability to integrate
conservation law into these bounds and produce better bounds based
on the fact that only evolution that is consistent with the conservation
laws has to be taken into account.

Strikingly, the passivity deformation framework simultaneously solves
these four cardinal deficiencies. However, we presently do not claim
that these solutions are unique or optimal. Nevertheless, passivity
deformation significantly improves the current thermodynamic predictive
power. In a companion paper to the present one \citep{uzdin2019IBMexp},
we have utilized the IBM platform mentioned above to experimentally
demonstrated that passivity deformation can produce bounds that outperform
both the second law and the global passivity inequalities. This demonstration
provides a positive indication that this framework is relevant for
modern experiments.

Recent years have seen important developments in our understanding
of the thermodynamics of small quantum systems. Stochastic thermodynamics
allows one to assign thermodynamic quantities such as heat or work
to a single trajectory of a colloidal particle or a molecular motor
\citep{Seifert2012StochasticReview,sekimotoStochEnergBook}. Quantum
thermodynamics aims to identify the thermodynamic role of purely quantum
effects such as coherence, measurement back action, or entanglement
\citep{Goold2015review,SaiJanetReview,RobReview2017,merali2017NatureNewsRev}.
Despite their success, both approaches are not well suited for the
goal discussed above due to their focus on the energy (work, heat,
and their fluctuations) as the observable of interest.

Our paper is structured as follows. After reviewing the notions of
passive operators, and global passivity, in Sec. II, Sec. III describes
the essence of the passivity deformation method. In addition to studying
the tightness of the new bounds and its intriguing physical meaning,
it is also shown how conservation laws can be integrated and yield
even better bounds. The section ends with several illustrative examples.
In Sec. IV we introduce an intuitive graphical representation of our
framework, which is exploited for deriving several useful bounds and
insights without doing explicit calculations. For example, we obtain
a more refined bound on information erasure compared to the Landauer
bound. We then use our framework to resolve the ultra-cold catastrophe
of the second law. At the end of this section, it is shown that some
non-thermal and potentially correlated environments can be treated
on the same footing as thermal uncorrelated environments, where the
deviation from thermal initial conditions reduces to using in new
effective temperatures in the familiar second law. In Sec. V passivity
deformation is utilized to study coarse-graining within the framework
of passivity, and also to unravel a hierarchal structure between the
second law and majorization condition. We conclude in Sec. VI.

\section{\label{sec: Formulation-of-global}passive states, passive operators,
and global passivity}

\subsection*{Passive states}

Passive states (passive density matrices) were introduced for studying
how much work can be extracted from an isolated system by using external
forces \citep{pusz78,lenard1978Gibbs,AllahverdyanErgotropy,MartiWorkCorr,WolgangSqueezedErgotropy}.
Mathematically, this requires finding the unitary transformation that
brings the system to the lowest energy. Crucially, the notion of passivity
is not limited to studies of energy changes. It can be applied to
other observables as well \citep{GlobalPassivity}. Consider a Hermitian
operator $A$, and a system whose initial state is described by a
density matrix $\rho_{0}$. $A$ may be the Hamiltonian of the whole
system, a subsystem, or may also describe other observables, such
as angular momentum, or projection operators onto specific subspaces.
One can then ask what is the minimal value of $\left\langle A\right\rangle =tr(\rho A)$
that is reachable from the initial state by a unitary transformation.
The state achieved by this optimal unitary is called ``passive state''
$\rho_{A\:pass}$ (with respect to the operator $A$). By construction
one obtains the inequality
\begin{equation}
\left\langle A\right\rangle \ge\underset{\text{all \ensuremath{U}}}{\min}tr(U\rho U^{\dagger}A)=tr(\rho_{A\:pass}A)\label{eq: min val pass-1}
\end{equation}
that holds for all unitaries $U$. This is just a definition, but
$\rho_{A\:pass}$ has an explicit expression. The operator $A$ can
be written in terms of its eigenvalues and eigenvectors, $A=\sum_{i}a_{i}\left|a_{i}\right\rangle \left\langle a_{i}\right|$,
with $a_{i+1}\geq a_{i}$. A general initial density matrix has the
form $\rho_{0}=\sum_{i}r_{i}\left|r_{i}\right\rangle \left\langle r_{i}\right|$.
A density matrix that is passive with respect to the operator $A$
will then have the form 
\begin{equation}
\rho_{A\:pass}=\sum_{i}r_{i}\ketbra{a_{i}}{a_{i}},\label{eq: pass state def-1}
\end{equation}
with $r_{i+1}\leq r_{i}$. Thus the optimal unitary is simply $U_{opt}=\ketbra{a_{i}}{r_{i}}$.
The ordering of $r_{i}$ with respect $a_{i}$ is crucial for passivity.
See proof in \citep{AllahverdyanErgotropy}. The conditions for passivity
(\ref{eq: pass state def-1}) can also be recasted as 
\begin{equation}
\left[A,\rho_{A~pass}\right]=0,~~~\left\langle a_{i+1}\right|\rho_{A~pass}\left|a_{i+1}\right\rangle \geq\left\langle a_{i}\right|\rho_{A~pass}\left|a_{i}\right\rangle .\label{eq: pass state def B}
\end{equation}
The definition of passivity given above is valid for any unitary matrix.
In particular, it can be used for the evolution operator of a setup
that includes all the elements that interact with each other. Thus,
if a setup was prepared in an initial state that is passive with respect
to an operator $A$, i.e. it is already has the minimal value of $\left\langle A\right\rangle $,
then any subsequent unitary evolution must satisfy the inequality
\begin{equation}
\Delta\left\langle A\right\rangle =tr(\rho_{f}A)-tr(\rho_{0}A)\ge0,\label{eq: pass initial state A-1}
\end{equation}
for $\rho_{0}=\rho_{A~pass}$. Moreover, by linearity, this inequality
also holds if the evolution is described by a mixture of unitaries.
\begin{equation}
\rho_{f}=\sum_{k}p_{k}U_{k}\rho_{0}U_{k}^{\dagger}.\label{eq: mix uni 2}
\end{equation}
Simply put, starting with the minimal value obtained by unitaries,
the expectation value can only grow with respect to its initial value.

\subsection{Passive operators}

\begin{figure}
\includegraphics[width=8.6cm]{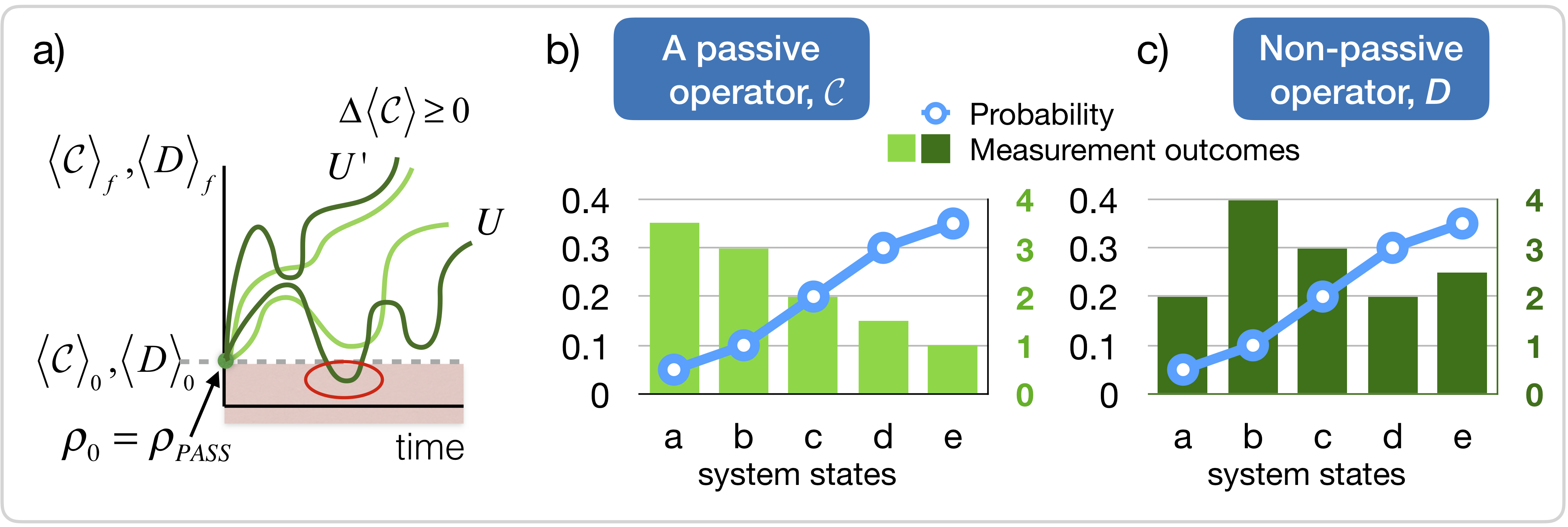}

\caption{\label{fig: pass op}(a) The expectation values of a passive operator
$\protect\mc C$ (light-green) must increase under any unitary operation
with respect to the initial value. The dark green curves show the
expectation values of a non-passive operator $D$. (b-c) Graphical
representation of passive/non passive operator in the basis of increasing
probabilities.}
\end{figure}

Passivity is not a property of the density matrix alone but a relation
between a density matrix and an observable (an operator). A given
initial state $\rho_{0}$, may be non passive with respect to the
Hamiltonian, but passive with respect to other operators. Thus, one
can use the definition of passivity to distinguish between passive
and non passive operators for a given density matrix. More explicitly,
writing $\rho=\sum r'_{i}\ketbra{r'_{i}}{r'_{i}}$ with $r'_{i+1}\ge r'_{i}$
(note the different ordering compared to \ref{eq: pass state def-1}),
a passive operator $A_{\rho_{0}\,pass}$ with respect to $\rho_{0}$
satisfies

\begin{equation}
\left[A_{\rho_{0}\,pass},\rho_{0}\right]=0,~~~\left\langle r'_{i+1}\right|\rho_{A~pass}\left|r'_{i+1}\right\rangle \geq\left\langle r'_{i}\right|\rho_{A~pass}\left|r'_{i}\right\rangle .\label{eq: pass op def B}
\end{equation}
Such operators satisfy
\begin{align}
\Delta\left\langle A_{\rho_{0}\,pass}\right\rangle  & \ge0,\label{eq: pass initial state pass op}
\end{align}
for any mixture of unitaries {[}Eq. (\ref{eq: mix uni 2}){]}. Fig.
\ref{fig: pass op} depicts an example of an initial density matrix
$\rho_{0}$ (dashed-dot curve) and two operators $C$ and $D$ (bars).
Both operators commute with $\rho_{0}$ so $\rho_{0},C,D$ can all
be conveniently plotted using the eigenvectors of $\rho_{0}$. Each
tick in the X axis of Figs. \ref{fig: pass op}a and \ref{fig: pass op}b
matches one of these eigenvectors, ordered so that the eigenvalues
of $\rho_{0}$ are monotonically increasing. The Y axis depicts the
eigenvalues of $\rho_{0}$ and of the operators. In this plot globally
passive operators are monotonically decreasing. Hence operator $C$
is passive while operator $D$ is not. If $\rho_{0}$ has a degeneracy,
the order of the degenerate states on the x axis can be chosen at
will. Thus, if the bar plot is increasing only in a degenerate subspace
of $\rho_{0}$, it is still passive. Alternatively, it is possible
to permute the degenerate states (their order was arbitrary to begin
with) so that plot becomes monotonically decreasing.

When the setup undergoes various unitary evolutions (green curves
in Fig. \ref{fig: pass op}c), the expectation value of a passive
operator will never go below its initial value (red zone). It contrast,
for a non passive operator (dark green curve) there is always a unitary
that reduces the expectation value below its initial value (i.e. enters
the red zone).

Importantly, given an initial state $\rho_{0}$ there are many operators
that satisfy passivity relations with $\rho_{0}$. Consequently, the
inequality (\ref{eq: pass initial state pass op}) applies for many
different operators. The pertinent questions which we address next
are i) whether these operators can be systematically constructed and
ii) whether one can find passive operators with an interesting physical
meaning.

\subsection{Global passivity}

In our previous work \citep{GlobalPassivity} we used $\rho_{0}$
itself to construct a passive operator with respect to the initial
density matrix,
\begin{equation}
\mc B=-\ln\rho_{0}^{tot},\label{eq: B0 def}
\end{equation}
where, crucially, $\rho_{0}^{tot}$ is the density matrix of the \emph{whole
setup} including both the system (if there is one) and the environments.
Note that by definition the operator $\mc B$ does not change in time
i.e. $\left\langle \mc B\right\rangle _{t}=tr[\rho_{t}^{tot}(-\ln\rho_{0}^{tot})]$.
The passivity of $\mc B$ with respect to $\rho_{0}$ immediately
follows from $\rho_{0}=e^{-\mc B}$. Consequently, it holds that
\begin{equation}
\triangle\left\langle \mc B\right\rangle \ge0\label{eq: B pass ineq}
\end{equation}
for any mixture of unitaries. The quantum evolution of a microscopic
system may not be unitary if it is not isolated from its environment.
Nevertheless, the evolution of both the system and its nearby environments
can be viewed as unitary if they are sufficiently isolated from the
rest of the universe. In such cases, any process can be modeled as
a unitary that acts on both the system and its local environment.
The approach we present here makes this assumption. We use the term
global passivity to highlight the fact that the processes we consider
involve unitary evolution that acts on both system and its local environments,
in contrast to the standard notion of passivity where one subsystem
is driven by an external field.

Without a meaningful physical interpretation, the inequality (\ref{eq: B pass ineq})
is just a mathematical result. Our previous work starts by pointing
out a clear connection between global passivity and the second law.
Let $\{\beta_{k},H_{k}\}$ describe the inverse temperatures and Hamiltonians
of a set of subsystems that act as environments. These potentially
microscopic environments are termed 'microbaths'. In contrast to large
baths, they may strongly deviate from thermal equilibrium when interacting
with each other or with some external forces. Hence, their temperature
refers only to their initial state. We considered a setup where several
microbaths interact with each other (e.g. as in an absorption refrigerator
\citep{MarkAbsorptionReview,k272,maslennikov2019quantum,MitchisonHuber2015CoherenceAssitedCooling,Correa2014EnhancedSciRep,palao13,mitchison2016realising}).
Such a setup can also describe heat engines and power refrigerators.
Since $\rho_{0}=\exp(\sum\beta_{k}\Delta\left\langle H_{k}\right\rangle )/Z$
($Z$ is a normalization factor), the global passivity of $\mc B$
(\ref{eq: B pass ineq}) yields
\begin{equation}
\sum_{k}\beta_{k}\Delta\left\langle H_{k}\right\rangle \ge0,\label{eq: GP microbaths}
\end{equation}
which is the Clausius inequality formulation of the second law for
microscopic setups with initially thermal subsystems. For completeness,
the full form of the Clausius inequality which includes a system that
starts in an arbitrary initial state is
\begin{equation}
\Delta S_{sys}+\sum_{k}\beta_{k}\Delta\left\langle H_{k}\right\rangle \ge0.\label{eq: the full CI}
\end{equation}
Here $\Delta S_{sys}$ is the change in the von Neumann entropy of
the system. Our goal is to obtain inequalities that constrain expectation
values of observables. Quantities such as $\Delta S_{sys}$ are considerably
harder to obtain experimentally, so we wish to avoid their appearance
in the inequalities when possible. In addition, (\ref{eq: the full CI})
reduces to (\ref{eq: GP microbaths}) under a periodic evolution of
the system.

We emphasize once again that $\beta_{k}$ in (\ref{eq: GP microbaths})
and (\ref{eq: the full CI}) refers only to the initial state of the
environments. To identify $\Delta\left\langle H_{k}\right\rangle $
with the change in the average energy of the $k$-th microbath, it
is essential that the Hamiltonian of the microbath at the end of the
process is equal to the Hamiltonian at the beginning of the process.
Furthermore, in (\ref{eq: GP microbaths}) and (\ref{eq: the full CI})
the terms $\Delta\left\langle H_{k}\right\rangle $ are not automatically
identified as 'heat'. Although one can do so, there are other legitimate
alternatives (see \citep{Wolgang2018passivityCI} and Sec. 28.3.6
or III.E in \citep{MySecondLawReview}). Ultimately, (\ref{eq: GP microbaths})
refers to average energy changes in the microbaths and is independent
of how heat and work are defined.

While Eqs. (\ref{eq: GP microbaths}) and (\ref{eq: the full CI})
resemble the familiar second law of thermodynamics, they deviate from
the classical thermodynamics result in several important aspects:
i) the microbaths can be small, and may substantially deviate from
their initial thermal state during the process; ii) the dynamics may
create entanglement and correlations between different subsystems;
iii) thermal relaxation with ideal heat baths are not included or
assumed; and iv) work can be done during the process, but some of
it may be done on the microbaths and not only on the system of interest.

The example above shows both a systematic construction of a passive
operator and a clear thermodynamic context (the second law). Yet,
this example adds nothing new on top of the known second law which
can easily be obtained from an information-based approach \citep{PeresBook,Esposito2010SecLaw,Goold2015review,Sagawa2012second,MySecondLawReview}.
The added value of the global passivity framework manifests when constructing
additional globally passive operators. In particular, in \citep{GlobalPassivity}
it was shown that $\mc B^{\alpha}$ is also passive with respect to
$\rho_{0}^{tot}$ for any $\alpha>0$ (the more general form is $\text{sgn}(\alpha)\mc B^{\alpha}$
for any real $\alpha$). Crucially, in several cases we found that
the resulting inequalities contain useful information that is not
included in Eqs. (\ref{eq: GP microbaths}) or (\ref{eq: the full CI}).
The added value of these inequalities has been recently experimentally
demonstrated in the IBM quantum processor platform.

As shown in Appendix II global passivity can be formulated as a binary
relation based on a matrix ordering function. The conditions for this
binary relation to become an equivalence relation are discussed as
well. While, for clarity, the paper is written in the conventional
formalism of passivity (\ref{eq: pass initial state A-1}) and (\ref{eq: pass op def B}),
the formalism in Appendix II is highly useful when exploring consequences
of passivity.

The global passivity approach described above is well suited for processes
where a collection of quantum systems was prepared in a known initial
density matrix and is then driven. The resulting inequalities tell
us what can not be achieved in any subsequent evolution. Yet the approach
has the same limitations the second law has: 1) it provide no useful
input on fine-grained observables and exotic heat machines that use
thermodynamic resources to perform non-thermodynamic tasks; 2) bounds
constructed from $\rho_{0}^{tot}$ using global passivity, cannot
be saturated when the environments are small; 3) The global passivity
bounds also suffer from the ultra-cold catastrophe.

\section{\label{sec: Introducing PD}the passivity deformation approach}

In what follows we present a new approach that overcomes the above-mentioned
limitations. Consider an observable of interest $A$ that satisfies
$[A,\rho_{0}^{tot}]=0$. If $A$ is passive with respect to  $\rho_{0}^{tot}$
then $\triangle\left\langle A\right\rangle \geq0.$ A more interesting
case is when $A$ is not globally passive. In such cases an inequality
can be constructed by starting from a passive operator, for instance
$\mathcal{B}=-\ln\rho_{0}^{tot}$, and defining the operator
\begin{equation}
B\left(\xi\right)=\mathcal{B}+\xi A.\label{eq: B xsi A}
\end{equation}
where $B\left(0\right)$ is globally passive by construction. It is
expected that there is a finite range of $\xi$ values for which $B\left(\xi\right)$
is also globally passive and therefore satisfies $\triangle\left\langle B\left(\xi\right)\right\rangle \geq0$.
As explained later, for discrete and finite systems there is always
$\xi\neq0$ for which $B(\xi)$ is globally passive.

\begin{figure}
\includegraphics[width=8.6cm]{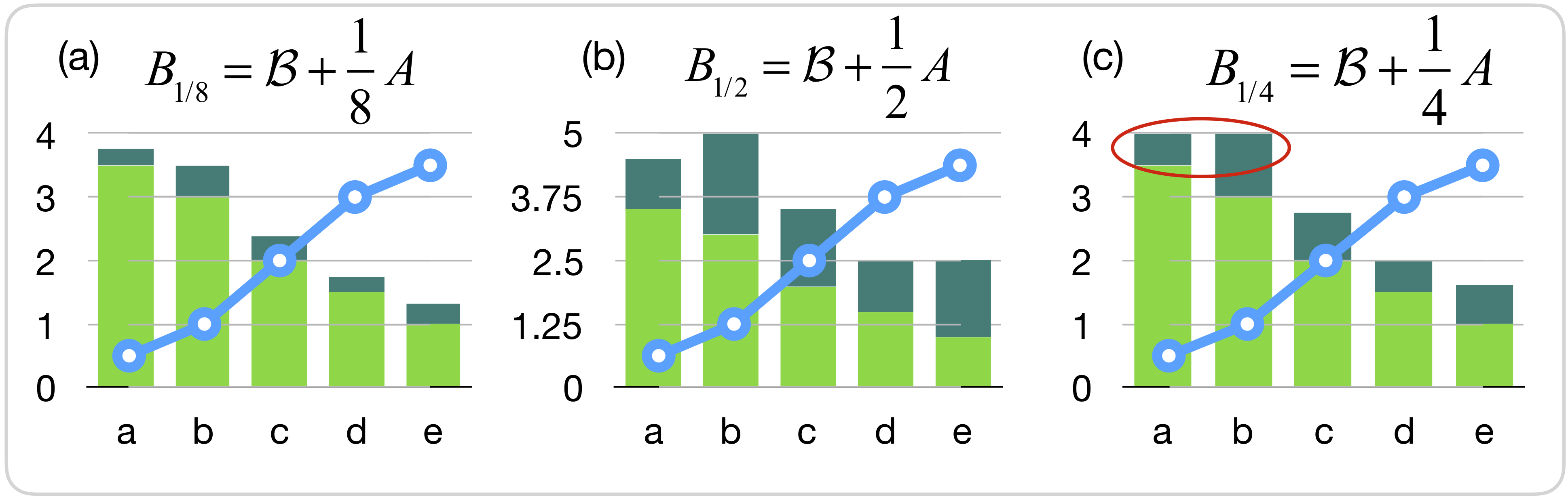}

\caption{\label{fig: pass const}Passivity deformation - when adding a small
amount of a non-passive operator (dark green) to a passive operator
(green) the combined operator created by the sum (depicted by the
total height) may still be passive as shown in (a). If the fraction
of the non-passive part is too large, passivity is lost (non-monotonically
decreasing). At some critical value in between, a new degeneracy forms
(c).}
\end{figure}

Which values of $\xi$ should be used? Tighter, and therefore more
restrictive and informative inequalities for $\left\langle A\right\rangle $
are obtained when $\left|\xi\right|$ is as large as possible. However,
at some point, some of the eigenvalues of $B\left(\xi\right)$ become
degenerate due to the change of $\xi$. Degeneracies inherited form
$B\left(0\right)$ are irrelevant at this point - only the ones emerging
from increasing $\left|\xi\right|$. At this critical value of $\xi$
the ordering of the operator changes and it stops being globally passive.
This is illustrated in Fig. \ref{fig: pass const}.

The condition for degeneracy between consecutive eigenvalues of $B\left(\xi\right)$
is $\lambda_{k+1}^{(\mc B)}+\xi\lambda_{k+1}^{(A)}=\lambda_{k}^{(\mc B)}+\xi\lambda_{k}^{(A)}$.
Let us define $\xi_{k}=\left(\lambda_{k+1}^{(\mc B)}-\lambda_{k}^{(\mc B)}\right)/\left(\lambda_{k}^{(A)}-\lambda_{k+1}^{(A)}\right),$
where $k$ values that nullify the numerator or denominator are excluded
since degeneracies in either $A$ or $\mc B$ do not affect the relative
ordering of the two operators. Using $\xi_{k}$ we define
\begin{align}
\xi_{+} & =\min(\xi_{k}>0),\label{eq: xi pm expression-1}\\
\xi_{-} & =-\min(-\xi_{k}>0).\label{eq: xi m expression-1}
\end{align}
The operator $B\left(\xi\right)$ is then passive in the range $\xi_{-}\leq\xi\leq\xi_{+}.$
Since $\xi_{k}$ cannot take the value zero (due to the exclusion
of $k$ values in the definition of $\xi_{k}$) there is a nontrivial
$\xi$ for which $B\left(\xi\right)$ is globally passive. In processes
for which $\Delta\left\langle A\right\rangle >0$ the tightest and
most informative inequality is found by using $\xi_{-}<0$, which
results in the inequality
\begin{align}
\Delta\left\langle A\right\rangle  & \le\frac{1}{(-\xi_{-})}\Delta\left\langle \mc B\right\rangle .\label{eq: PC const +change}
\end{align}
Similarly in processes in which $\Delta\left\langle A\right\rangle <0$
one should use $\xi_{+}$, giving
\begin{equation}
-\Delta\left\langle A\right\rangle \le\frac{1}{\xi_{+}}\Delta\left\langle \mc B\right\rangle .\label{eq: PC const -}
\end{equation}

The two inequalities (\ref{eq: PC const +change}) and (\ref{eq: PC const -})
should be viewed as restricting the change of an observable $A$ when
compared to the change on another, passive observable. In many of
the examples we present, $\mathcal{B}=-\ln\rho_{0}^{tot}$ will describe
a collection of microbaths so that $\mc B=\sum_{k}\beta_{k}H_{k}.$
Therefore, $\triangle\left\langle \mathcal{B}\right\rangle $ contains
information about energy changes of subsystems during the process.
In contrast, $A$ may describe a fine-grained non-thermal property,
for instance, the probability to be in a specific state. The inequalities
(\ref{eq: PC const +change}) and (\ref{eq: PC const -}) then describe
how changes in the expectation values of $A$ are restricted by the
subsystems energy changes. If $A$ happens to be globally passive
then $\Delta\left\langle A\right\rangle \ge0$ and (\ref{eq: PC const +change})
sets an upper bound on the change in $\left\langle A\right\rangle $.

These inequalities include setup-specific information through $\xi_{\pm}$,
which depend on the eigenvalues of $A$ and $\mathcal{B}$. Thus,
the method presented here allows to obtain tighter and more informative
inequalities compared to approaches that do not exploit such information
(such as the standard Clausius inequality and global passivity).

\subsection{\label{subsec: conserved quantities}Thermodynamic bounds in the
presence of conserved quantities}

Another major advantage of this scheme appears in the presence of
conserved quantities. In some cases the allowed unitary evolution
only couples states in subsets of the Hilbert space, while states
in different subsets are not coupled. As a result, the probability
to be in each decoupled subspace is conserved. A simple physical example
is energy conserving interactions between various elements in the
setup. Each energy shell of the setup is a closed manifold of states
that does not interact with the other energy shells. This restricted
dynamics may deny the possibility of executing the unitary that saturates
a certain global passivity inequality. Thus, this bound cannot be
attained due to the conservation law.

In passivity deformation, this can be avoided. Better and tight bounds
can be obtained by treating each manifold separately. For example
for $\xi_{+}$ instead of $\min(\xi_{k}>0)$ we calculate 
\begin{equation}
\xi_{+}^{\text{int}}=\min\{\min(\xi_{k\in\{l_{1}\}}>0),\min(\xi_{k\in\{l_{2}\}}>0),..\}\label{eq: xi conservation}
\end{equation}
where due to the conservation law the set of states $\{l_{i}\}$ never
interacts with the set $\{l_{j\neq i}\}$. This is a \emph{major advantage
of the passivity deformation framework as it allows the integration
of conservation laws} directly into the second law-like inequalities
obtained by global passivity.

\begin{figure}
\includegraphics[width=8.6cm]{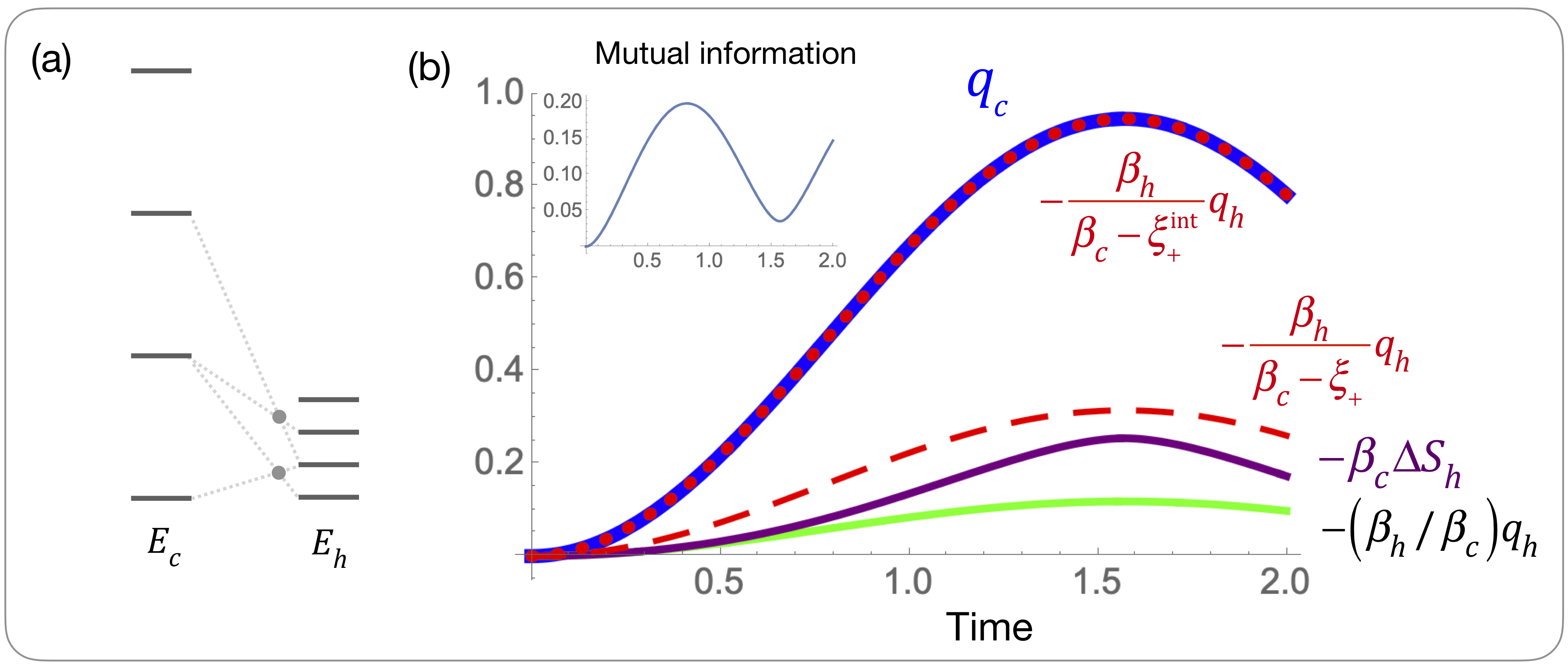}

\caption{\label{fig: two 4 level}(a) Two four-level systems (microbaths) with
initial inverse temperatures $\beta_{c}$,$\beta_{h}$ interact via
creation-annihilation terms. (b) The blue curve shows the actual change
of the average energy of the cold microbath $q_{c}=\Delta\left\langle H_{c}\right\rangle $.
The green line depicts the bound obtained from the second law (\ref{eq: GP microbaths}).
The entropy-based second law bound from Eq. (\ref{eq: the full CI})
is shown in purple. The basic passivity deformation prediction in
terms of heat (dashed-red) perform s better than both forms of the
second law. The passivity deformation bound that uses also the information
about conserved quantities (dotted-red) is \emph{tight }(analytically).
The inset shows that the mutual information (correlation) is not zero
in the process. This emphasizes two of the main strengths of the present
framework: i) it is tight in the presence of correlation and ii) it
can take conservation laws into account and produce tight bounds despite
the conservation constraint.}
\end{figure}

As an example, consider two four-level microbaths, initially at inverse
temperatures $\beta_{c}=2$ and $\beta_{h}=1$. The energy levels
are $E_{c}=\{0,4,8,12\}$ and $E_{h}=\{0,1,2,3\}$. The initial state
of the system is therefore described by the density matrix $\rho_{0}^{tot}=\exp\left(-\beta_{c}H_{c}\right)/Z_{c}\otimes\exp\left(-\beta_{h}H_{h}\right)/Z_{h}$.
The unitary evolution is generated by creation-annihilation interactions
that couples only the first three levels $H_{int}=a_{12}b_{12}^{\dagger}+a_{23}b_{23}^{\dagger}+h.c.$
where $a_{ij}$ ($b$) is the annihilation operator $\ketbra ij$
in the cold (hot) microbath. (see Fig. \ref{fig: two 4 level}a).
As a result, there are several conserved quantities, e.g. the population
of the fourth level in each microbath is conserved. Note that the
average energy is not conserved since we chose different energy spacings
in the two microbaths. Thus the microbath exchange energy not only
with each other but also with the external field that generates $H_{int}$.
At the end of the evolution the interaction is switched off.

In Fig. \ref{fig: two 4 level}b, the bounds on $\Delta\left\langle H_{c}\right\rangle $
by different forms of the second law and passivity deformation are
compared to the numerically calculated value of $\Delta\left\langle H_{c}\right\rangle $
(denoted by $q_{c}$ in the figure for brevity). The blue curve stands
for the exact value of $\Delta\left\langle H_{c}\right\rangle $.
The green line shows the bound imposed by the (\ref{eq: GP microbaths})
form of the second law, and purple line shows the bound imposed by
(\ref{eq: the full CI}). The dashed-red line represents the passivity
deformation prediction based on (\ref{eq: xi pm expression-1}) where
$\xi_{+}=5/8\beta_{c}$. Exploiting the conserved quantities in this
specific interaction we use (\ref{eq: xi conservation}) and get $\xi_{+}^{\text{int}}=7/8\beta_{c}$.
This latter bound $\Delta\left\langle H_{c}\right\rangle \ge\frac{\beta_{h}}{\beta_{c}(1-7/8)}\Delta\left\langle H_{h}\right\rangle $,
is tight in this example as can be seen by the dotted-red line.

Remarkably the bound is tight although there is a significant correlation
between the two objects as shown in the inset of Fig. \ref{fig: two 4 level}b.
In \citep{EspositoArakiLeib} it was indicated that for large environments
the correlation is not the main mechanism that makes the second law
non tight. It is the deviation from equilibrium in the reservoir that
plays a major role. In this example both mechanisms are important.
Nevertheless, we see that bound with $\xi_{+}^{\text{int}}$ is tight
despite the deviation from equilibrium in the microbaths. %

\subsection{\label{subsec: sat-path-indep}Bound-saturating, path-independent
operations}

In the previous subsection, we derived a bound of the form $\triangle\left\langle B\left(\xi\right)\right\rangle \geq0$,
where $B\left(\xi\right)=\mathcal{B}+\xi A$ is the sum of a passive
and a non-passive operators, that was valid for a finite range of
$\xi$. It is of interest to understand the properties of processes
that saturate this bound. This can be further motivated by an analogy
with traditional thermodynamics. The Clausius inequality $\triangle S_{sys}+\sum_{k}\beta_{k}q_{k}\geq0$
is saturated by reversible processes. Reversible processes keep the
reservoirs infinitesimally close to equilibrium and do not generate
classical correlations and/or entanglement between the system and
reservoirs. Reversibility constrains the process, and enable to exactly
express changes in one element of the setup in term of changes in
the others. For instance, in a reversible process, the change in the
entropy of the system (e.g. engine core) is fully determined by the
changes in the energy of the microbaths (heat) $q_{k}$: $\triangle S_{sys}^{R}=-\sum_{k}\beta_{k}q_{k}^{R}$.
Two very different reversible processes that have the same $\triangle S_{sys}^{R}$
(they can even involve completely different levels of the system)
will have the same ``weighted heat'' $\sum_{k}\beta_{k}q_{k}^{R}$.
In particular, in the special case of a single bath $\triangle S_{sys}^{R}$
fixes the heat exchanged with the bath and makes it path-independent.
If the two processes also have the same energy change in the system,
then the work becomes path independent as well since $W=\Delta F=\Delta(S_{sys}-\beta\text{\ensuremath{\left\langle H_{sys}\right\rangle }})$.
That is, by specifying the initial density matrix of the system alone,
it is possible to uniquely determine how much work and heat were invested
to make this change in a reversible manner.

In analogy to the path-independence associated with saturating the
CI, we ask what are the processes that saturate the bound $\Delta\left\langle B\left(\xi\right)\right\rangle =0$
. A trivial way to saturate this bound, is to apply processes that
leave the initial density matrix unchanged. This does not necessarily
imply that $U=I$. If there are degeneracies in $\rho_{0}^{tot}$,
then there is a family of unitaries that only mix states within each
degenerate subspace, and for any mixture of such unitaries $\rho_{f}^{tot}=\sum_{k}p_{k}U_{k}\rho_{0}^{tot}U_{k}^{\dagger}=\rho_{0}^{tot}$.
Such operations trivially lead to $\triangle\left\langle B\left(\xi\right)\right\rangle =0$
(as for any other expectation value). Thus, these trivial degeneracies
are of no interest to us.

A non-trivial way of saturating the bound can be found when the operator
$B\left(\xi\right)=\mathcal{B}+\xi A$ has degeneracies which are
different from the trivial degeneracies of $\mc B$. Interestingly,
\emph{these degeneracies are guaranteed to appear} at $\xi=\xi_{\pm}.$
Such degeneracies allow for non trivial processes that redistribute
population between these degenerate states of $B\left(\xi\right)$,
while keeping expectation value $\left\langle B\left(\xi\right)\right\rangle $
fixed (assuming no other operations take place). For a general mixture
of unitaries, we have the inequalities (\ref{eq: PC const -}), (\ref{eq: PC const +change}).
However, if we restrict the dynamics to be a mixture of unitaries
that only couple the states that become degenerate at $\xi_{\pm}$,
the inequality can be replaced by the equality 
\begin{equation}
\sum_{k}\beta_{k}q_{k}^{BSP}=-\xi_{\pm}\triangle\left\langle A\right\rangle _{BSP},\label{eq: BSP HR CI}
\end{equation}
where the index BSP indicates that only unitaries limited to this
$B\left(\xi_{\pm}\right)$ degenerate subspace are included. Crucially,
an interaction between nontrivially degenerates state leads to $\rho_{f}^{tot}\neq\rho_{0}^{tot}$
since these states are associated with different initial probabilities. 

The implications of (\ref{eq: BSP HR CI}) are similar to the familiar
reversible path-independence mentioned above: knowing that change
in $\triangle\left\langle A\right\rangle $ was created by a BSP,
fixes the value of $\sum_{k}\beta_{k}q_{k}$ regardless which BSP
was actually used. Note that the BSP for $\xi_{-}$ and $\xi_{+}$
are necessarily different from each other since otherwise there will
be two conflicting predictions on $\sum_{k}\beta_{k}q_{k}$.

We wish to draw the attention of the reader that since $\mc B_{\xi}$
and $A$ in (\ref{eq: BSP HR CI}) have different eigenvalue ordering
when written in the same basis, it follows that they cannot be minimized
at the same time (although they commute). Thus, there is a generic
tradeoff between saturating the $B_{\xi}$ bound and performance (minimizing
or maximizing $\left\langle A\right\rangle $). In contrast to the
familiar power-efficiency tradeoff in heat machines, the present tradeoff
has nothing to do with time and adiabaticity. It refers to the total
accumulated effect and time plays no role here.

While reversible operations seem more general and generic than system-specific
BSP, this is not the case in small setups. First, while BSP are guaranteed
to exist whenever $\xi_{-}$ or $\xi_{+}$ are different from zero,
reversible operations such as isotherms cannot be implemented in small
isolated setups. There are two reasons for this: 1) the microbaths
develop non-negligible correlation 2) due to their small heat capacity
the microbaths do not remain locally in equilibrium with their original
temperature. In contrast, the BSP are standard unitary operations
that can be implemented with suitable control fields. That is, in
microscopic setup the dynamics is thermodynamically irreversible (CI
not saturated, i.e., nonzero entropy production). \emph{Nevertheless,
the BSP saturates the passivity deformation bounds despite the thermodynamic
irreversibility indicated by the non-zero entropy production.}

In summary, by examining the structure of the passive operators of
the form $B(\xi_{\pm})$ and restricting the allowed unitaries to
those associated with the emerging $\xi_{\pm}$ degeneracies we identify
non trivial processes that saturate the passivity construction bounds.
The path-independence associated with the equality, bare some resemblance
to reversible processes. However, the reasons for the saturation of
the bound are quite different in both instances. Proximity to equilibrium
during the dynamics in one case, and a restriction to evolution in
a degenerate subspace in the other.

\subsection{Illustrative examples}

In this section, we present several examples that demonstrate how
to obtain inequalities with interesting physical interpretation using
the passivity deformation approach. These examples are intentionally
chosen to be elementary and involve just a few particles. We aim to
show how this method works in the simplest setups and what kind of
results it can provide. It is straightforward to apply it to larger
setups where the dynamics is highly non-trivial. An additional set
of examples will be presented later, after introducing a graphical
approach to passivity deformation.

\subsubsection{\label{subsec: X machine examp}Performance of an exotic heat machine}

Consider the machine described in Fig. 1, namely a two-spin system
that can be manipulated by interacting with a microbath. To keep the
plot simple and tractable we use a two-spin environment (microbath).
The goal of the setup is to implement a protocol of interaction with
the environment that will make the two spins of the system as correlated
as possible. Specifically, we wish the system spins to be in the $\ket{00}_{sys}$
state or in the $\ket{11}_{sys}$ state. The Hamiltonian of the setup
is $H=H_{sys}+H_{env}+H_{int}(t)$, where $H_{sys}=\omega\sigma_{1}^{z}+\omega\sigma_{2}^{z}$,
$H_{env}=\omega\sigma_{3}^{z}+\omega\sigma_{4}^{z}$, and we set $\omega=1$
for simplicity. $H_{int}(t)$ depends on the protocol used for the
correlation enhancement. Initially $H_{int}(0)=0$, and the initial
inverse temperature is $\beta=1/2$.

\begin{figure}
\includegraphics[width=8.6cm]{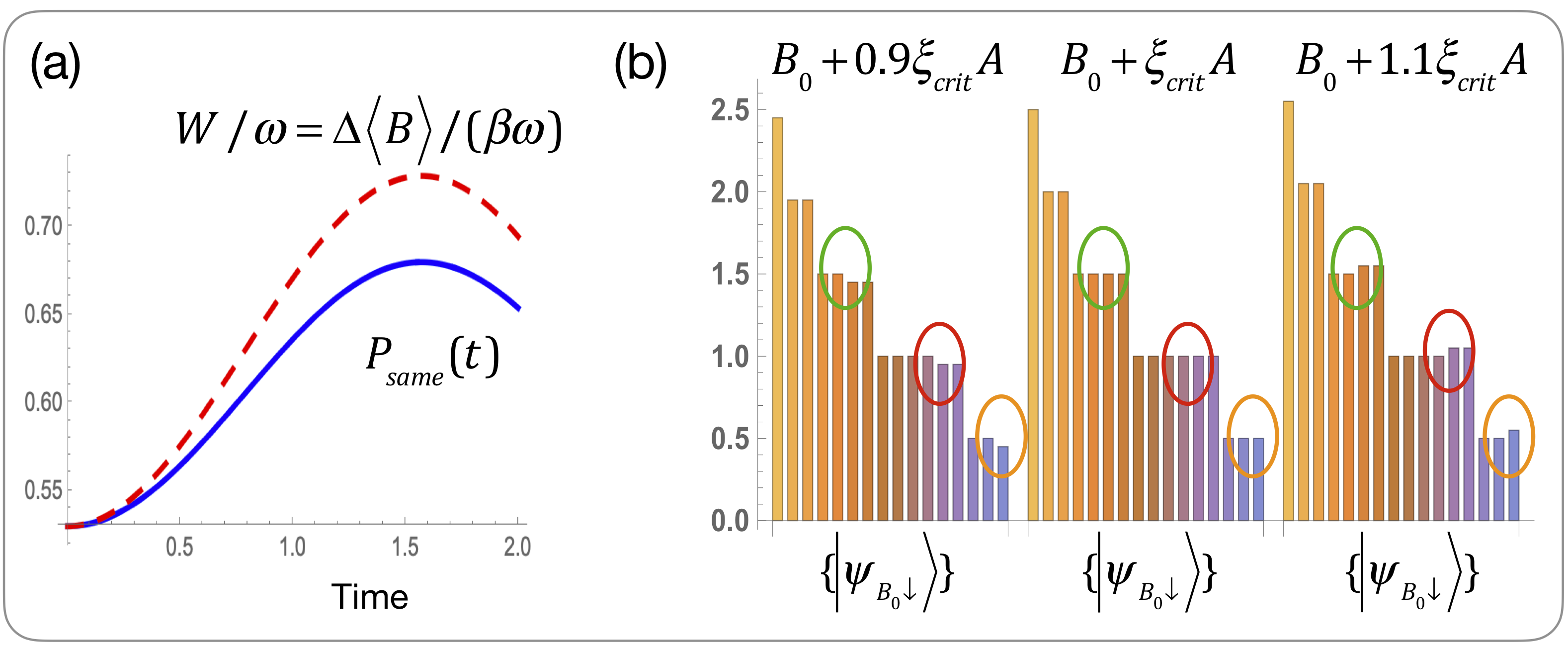}

\caption{\label{fig: corr machine bars}(a) The blue curve shows the change
in the correlation of the two spins in the machine described in Fig.
\ref{fig: X machine}. Passivity deformation predicts that the (scaled)
work (dashed-red) is always larger than the correlation creation (blue).
(b) The bar shows the emergence of degeneracies before, at, and after
the critical point $\xi_{-}$. Unitary operation between states in
the same ellipse exhibit path-independence behavior.}
\end{figure}

The probabilities of to be in a certain set of states is given by
expectation values of projection operator to this set of state. In
the present case the goal is to increase the expectation value of
the projector $A=\ketbra{00}{00}_{sys}+\ketbra{11}{11}_{sys}$, since
$P_{same}=P(11)+P(00)=tr(\rho A)=\left\langle A\right\rangle $. What
are the limitations on this class of processes? And which resources
must be invested to generate the desired output? Using the form $B_{\xi}=\mc B+\xi A$,
and employing Eq.(\ref{eq: PC const -}), we find that $\xi_{opt}=\beta$.
As a result, we obtain the bound
\begin{equation}
\Delta P_{same}\le W/\omega,\label{eq: corr X machine ineq}
\end{equation}
where $W=\Delta\left\langle H_{sys}+H_{env}\right\rangle $ is the
work done on the setup during the process. We conclude that our ability
to realign the system's spins is directly bounded by how much work
we invest. Note that this result holds for any $\beta$ and any $\omega$.
Using the method described in Appendix II we find the unitary that
maximize $\left\langle A\right\rangle $. Figure \ref{fig: corr machine bars}a
shows that in accordance to the passivity construction prediction,
$\Delta P_{same}\le W/\omega$, the work (normalized by $\omega$)
that has to be invested (red line) is larger than the increase in
the probability of the spins to be in the same orientation. As explained
at the end of Sec. \ref{subsec: sat-path-indep} the saturation of
the passivity deformation bounds typically conflicts with achieving
the maximal performance. This explains why the bound is not tight
in this example, as we have chosen the unitary that maximizes the
performance.

Figure \ref{fig: corr machine bars}b shows the emergence of new degeneracies
at $\xi=\beta$. If the dynamic is restricted to unitaries that mix
only states inside each ellipse, then it is guaranteed that (\ref{eq: corr X machine ineq})
will become equality. That is, the amount of work will not depend
on which BSP transformation we applied only on the change in the correlation
observable $\Delta P_{same}$, i.e. there is path-independence for
BSP operations.

The second law is not the correct tool for setting performance bounds
for such machines as it does not contain the changes in the fine-grained
observable $A$. This example illustrates the utility of our approach
to quantifying the performance limits of such exotic heat machines.

\subsubsection{\label{subsec: dephasing}Bounds on system-environment covariance
in dephasing dynamics}

To show a more quantum aspect of our approach we consider dephasing
dynamics. In our previous work \citep{GlobalPassivity} we obtained
a bound on the covariance between the coherence of a system and the
energy of its dephasing environment, i.e. $\left\langle \sigma_{x}H_{\mu b}\right\rangle -\left\langle \sigma_{x}\right\rangle \left\langle H_{\mu b}\right\rangle $.
The system was a spin with some initial coherence in the energy basis,
and it interacted with a three-spin microbath (environment) via an
interaction of the form $H_{int}=\sum_{j=1}^{3}\gamma_{j}\sigma_{z}^{sys}\otimes\sigma_{z}^{(i)}$.
Here $\gamma_{i}=\{0.7,0.5,0.3\}$ is a set of couplings that represents
the case where some environment spins are further away from the system.
The system Hamiltonian is $H_{sys}=\sigma_{z}^{sys}$ and the microbath
Hamiltonian is $H_{\mu b}=\sum_{j=1}^{3}\sigma_{z}^{(j)}$. The initial
state of the setup $\rho_{0}^{tot}=\exp(-\beta_{x}\sigma_{x}^{sys}-\beta H_{\mu b})/Z$
with $\beta_{x}=\beta=3$. From the initial density matrix we obtain
$\mathcal{B}=\beta_{x}\sigma_{x}^{sys}+\beta H_{\mu b}+\ln Z.$ 

\begin{figure}
\includegraphics[width=8.6cm]{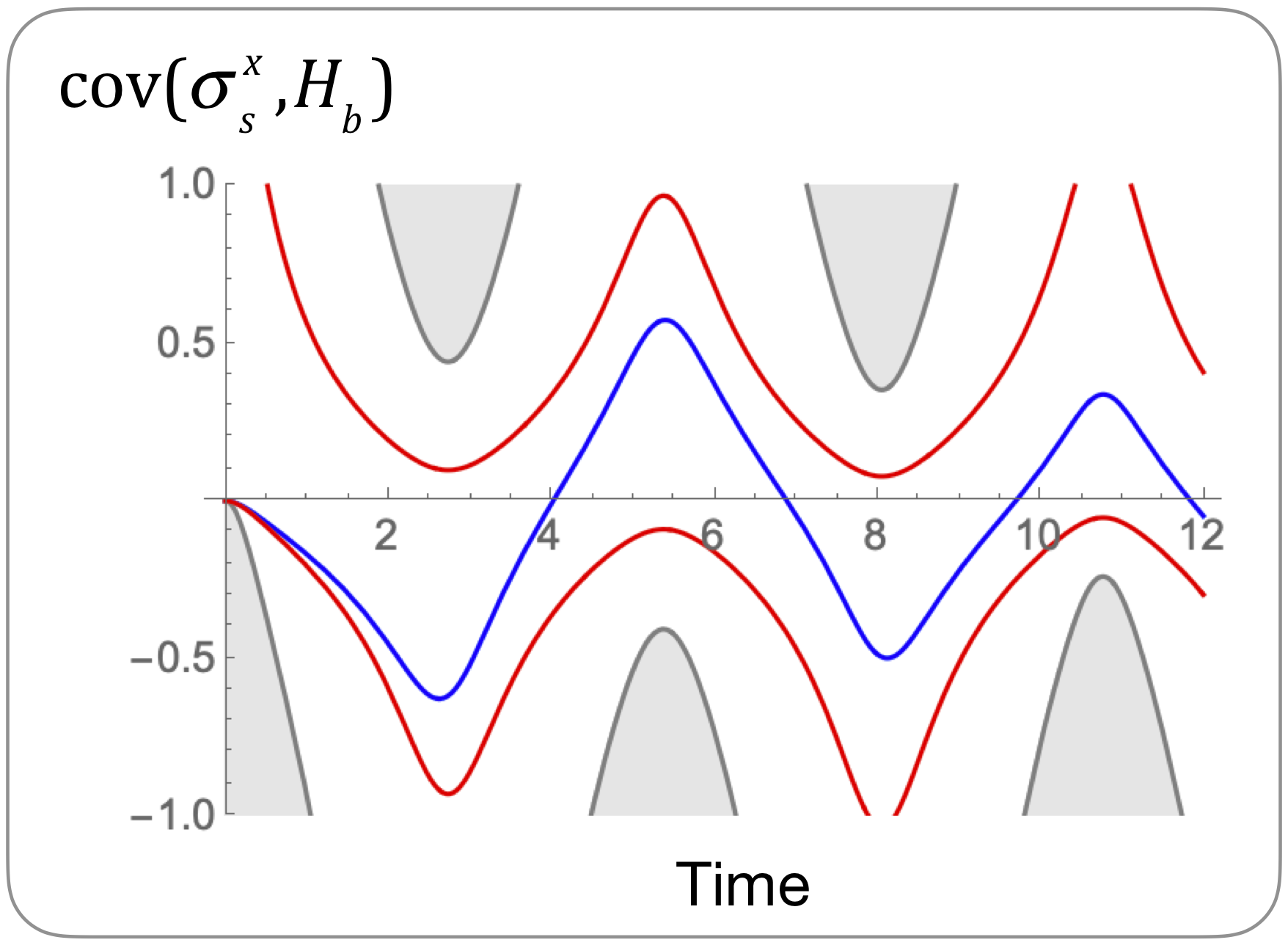}

\caption{\label{fig: Cov dephase}The dynamics of the normalized covariance
between the polarization of an initially coherent spin and the energy
of a dephasing environment composed of three thermal spins. The exact
dynamics is shown in blue and the gray areas are the forbidden zones
according to the global passivity bound $\Delta\left\langle B^{2}\right\rangle \ge0$.
Using the passivity construction framework we find significantly tighter
bounds (red).}
\end{figure}

In \citep{GlobalPassivity} we have used the global passivity of $\mathcal{B}^{2}$
to set a bound on $\left\langle \sigma_{x}^{sys}H_{\mu b}\right\rangle $.
Here we look for a tighter inequality by constructing $B\left(\xi\right)=\mc B^{2}+\xi\sigma_{x}^{sys}$
and studying for which $\xi$ values $B\left(\xi\right)$ is globally
passive. Figure \ref{fig: Cov dephase} shows in grey the bounds obtained
in \citep{GlobalPassivity} based on $\Delta\left\langle \mathcal{B}^{2}\right\rangle \ge0$,
and in red the passivity deformation bounds $\Delta\left\langle B\left(\xi_{-}\right)\right\rangle \ge0$
with $\xi_{-}=-9$. See \citep{GlobalPassivity} for the technique
used for deriving the upper bounds. Clearly, the passivity deformation
bound is closer to the actual covariance dynamics (blue) compared
to the global passivity bound.

\subsubsection{\label{subsec: Demon-detection}Demon detection through violation
of passivity deformation bounds}

One of the ways in which inequalities of the type derived here can
be useful is through their violation, which let us know that some
of the assumptions made on the dynamics of the setup must be broken.
In Ref. \citep{GlobalPassivity} we used this idea for detecting the
presence of Maxwell-like demons that may tamper with the dynamics.
The demons that were considered were too subtle to be detected by
the second law. Nevertheless, there were detectable by the violation
of some of the global passivity inequalities $\Delta\left\langle \mathcal{B}^{\alpha}\right\rangle \ge0$.
As it turns out, the family of inequalities $\Delta\left\langle \mathcal{B}^{\alpha}\right\rangle \ge0$
is not sensitive enough for detecting any demon. This raises the key
question regarding the existence of more refined thermodynamic tests
(inequalities) that can detect the subtle tampering of these ``lazy
demons''.

Consider two initially uncoupled microbaths at different temperatures
i.e. $\rho_{0}^{tot}=\exp\left(-\beta_{c}H_{c}\right)/Z_{c}\otimes\exp\left(-\beta_{h}H_{h}\right)/Z_{h}$.
In the absence of external work, the second law assures us that subsequent
evolution will result in energy transfer from the hot to the cold
microbath. If these two microbaths are well isolated from the rest
of the world and there are no demons the evolution is unitary $\rho_{f}^{tot}=U\rho_{0}^{tot}U^{\dagger}$.
For simplicity, we consider a demon that is operating on the setup
at the end of the unitary evolution. The demon applies feedback that
depends on the state of the setup, resulting in dynamics described
by a Kraus map $\tilde{\rho}_{f}^{tot}=\sum_{k}U_{k}\Pi_{k}\rho_{f}^{tot}\Pi_{k}U_{k}^{\dagger}$
where $\Pi_{k}$ are projectors to measurement outcome $k$. Passivity
based inequalities are not guaranteed to hold under such evolution.
There are two ways to make demon detection more challenging. The first
is to apply mild operations. That is, to use demon (feedback) operation
$U_{k}$ which are very close to the identity. The second option is
to apply the feedback with some probability $p$, and with probability
$1-p$ do nothing so that $\tilde{\rho}_{f}^{tot}=p\sum_{k}U_{k}\Pi_{k}\rho_{f}^{tot}\Pi_{k}U_{k}^{\dagger}+(1-p)\rho_{f}^{tot}$.
Consequently, for the same feedback operations $U_{k}$, $p=0$ corresponds
to demon-free evolution while $p=1$ is the standard Maxwell demon
(that violates the second law). A lazy demon is a demon with low enough
value of $p$ so that $\beta_{c}\Delta\left\langle H_{c}\right\rangle +\beta_{h}\Delta\left\langle H_{h}\right\rangle \ge0$
and the second law cannot be used to detect the demon.

In the lazy demon scenario considered in \citep{GlobalPassivity}
it was found that the optimal detection using the $\Delta\left\langle \mc B^{\alpha}\right\rangle $
family of inequalities, takes place at $\alpha\simeq2.56$ (this value
is not universal). This suggests that higher $\alpha$ value are not
necessarily better, and that a non integer value of $\alpha$ can
have a practical advantage. Most importantly, the standard second
law ($\alpha=1$) did not detect this demon.

We return to this example and test whether one can find even more
sensitive inequalities using the approach developed here. In the example
used in Ref. \citep{GlobalPassivity} both microbaths were two-spin
systems, with $H_{c}=\sigma_{z}^{(1)}+\sigma_{z}^{(2)}$ and $H_{h}=\sigma_{z}^{(3)}+\sigma_{z}^{(4)}$.
The initial inverse temperatures were chosen to be $\beta_{c}=2/3$
and $\beta_{h}=0.4$. To derive passivity deformation based inequalities,
we assume that the dynamics in the absence of a demon is unitary and
construct the operator $B_{\sigma_{z}^{(3)}}\left(\xi\right)=\beta_{c}H_{c}+\beta_{h}H_{h}+\xi\sigma_{z}^{(3)}$.
Using (\ref{eq: PC const +change}) and (\ref{eq: PC const -}) we
find that the range of $\xi$ values for which this operator is passive
is between $\xi_{-}\approx-0.266$ and $\xi_{+}\approx0.133$.

We now consider a process that involves evolution with an ``all to
all'' interaction between the spins $H_{I}=\sum_{i>j}\sigma_{+}^{(i)}\sigma_{-}^{(j)}+\sigma_{-}^{(i)}\sigma_{+}^{(j)}.$
After the evolution, the demon is awake with probability $p$. If
it is awake and the system is in the state $\left|1100\right\rangle $
it replaces it with the state $\left|0011\right\rangle $. In all
other cases, the demon does nothing. In this example, a demon that
operates 0.56 of the time or more will violate the Clausius inequality.
In \citet{GlobalPassivity} it was shown that a demon that operates
only 0.48 of the time (or more) will violate the inequality $\triangle\left\langle \mathcal{B}^{\alpha}\right\rangle \geq0$
(with $\alpha=2.56$) Crucially, using passivity deformation we find
that a demon that operates more than 0.289 of the time will violate
the inequality $\Delta\left\langle B_{\sigma_{z}^{(3)}}\left(\xi_{\pm}\right)\right\rangle \ge0$
derived here. Hence the passivity deformation approach leads to more
sensitive detection compared to global passivity.

This example, as well as the previous examples studied in this section,
demonstrate that the inequalities (\ref{eq: PC const +change}) and
(\ref{eq: PC const -}) are tighter, and therefore more informative
than Clausius inequality. Moreover, they are tight even in the presence
of small environments and correlation buildup. This comes at a certain
cost. To derive these new bounds, one has to exploit system-specific
information about the eigenvalues of relevant operators, and on the
initial state of the setup. This setup-dependence of $\xi$ is also
the reason that these bounds often surpass the prediction of the second
law - they exclude scenarios that cannot even happen in the setup
of interest (e.g. see the discussion on conserved quantities). In
many modern experimental setups such as superconducting qubits, trapped
ions, or optical lattices, the initial state and the Hamiltonian is
known. Hence, the inequalities studied here are well suited for the
description of quantum processes in such setups.

\section{\label{sec: PD-graphic}Additional insights from a graphical representation
of passivity deformation}

\begin{figure}
\includegraphics[width=8.6cm]{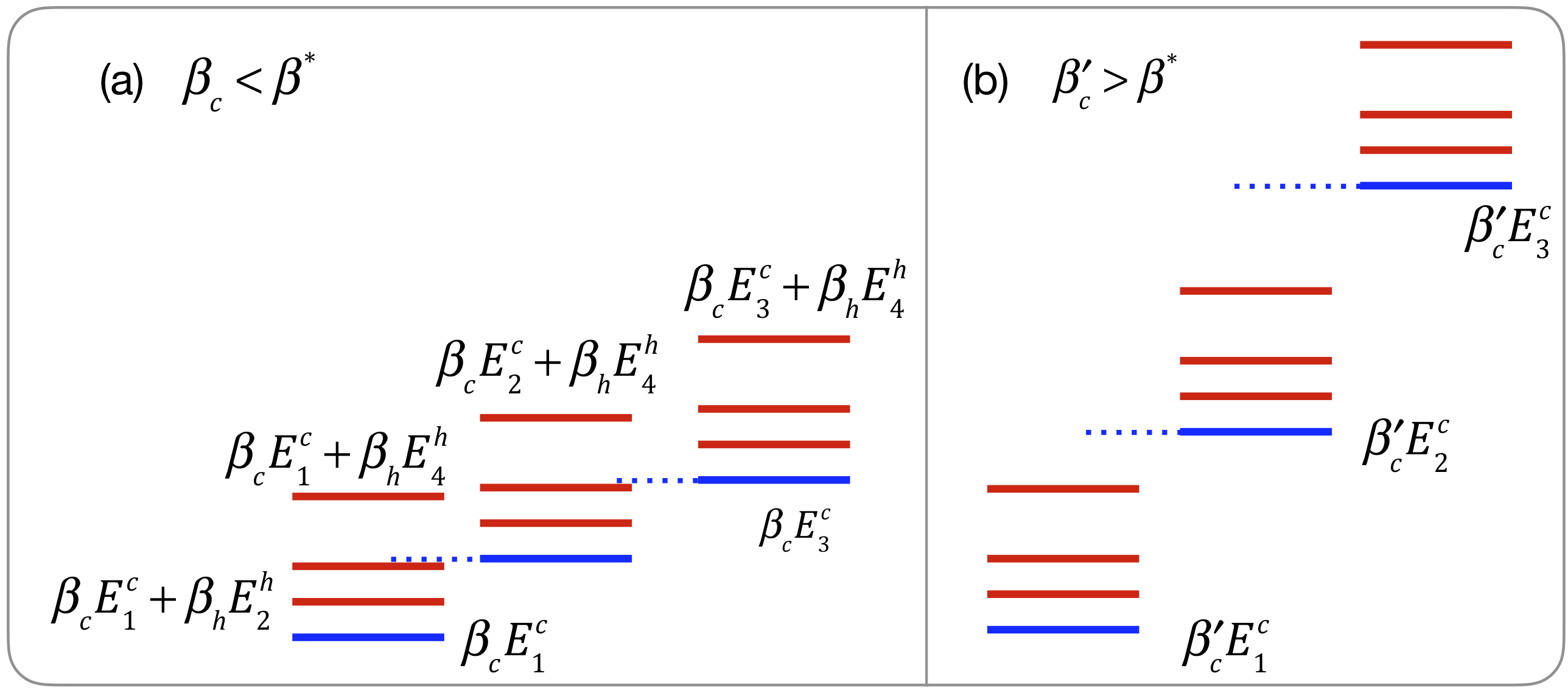}\caption{\label{fig: pass prod UC}Plotting the spectrum of a globally passive
operator of the form $\beta_{c}H_{c}+\beta_{h}H_{h}$. For high enough
temperature $T_{c}$ (low $\beta_{c}$) the hot manifolds overlap
(a) while for cold enough temperature (high $\beta_{c}$) they do
not.}
\end{figure}

\noindent In this section, we explore a more visual and intuitive
method of obtaining passivity deformation based inequalities. The
basic idea is to start with the globally passive operator $\mc B=-\ln\rho_{0}^{tot}$,
and deform it to a new operator $\mathcal{\widetilde{B}}$, by shifting
some of its levels (eigenvalues) using a set of rules that ensure
the resulting operator is globally passive. As a result, the subsequent
evolution will satisfy 
\begin{equation}
\triangle\left\langle \mathcal{\widetilde{B}}\right\rangle \geq0.
\end{equation}
Although later we will consider more complicated scenarios, lets us
start with the basic scenario of two uncorrelated microbaths for which
\begin{equation}
\mc B=\beta_{h}H_{h}\otimes I_{c}+I_{h}\otimes\beta_{c}H_{c},\label{eq:Btwomub}
\end{equation}
where we have explicitly denoted the identity operators in each subspace.
We also dropped an additive constant, arising from the normalization
of $\rho_{0}^{tot},$ that would not affect the resulting inequality.
Since the two terms commute the eigenvalues of $\mc B$ have the form
$\beta_{c}E_{\lambda}^{(c)}+\beta_{h}E_{\nu}^{(h)}$ with $E_{\lambda}^{(c)}$($E_{\nu}^{(h)}$)
denoting the eigenvalues of $H_{c}$($H_{h}$). It is most useful
to plot this spectrum using ``floors'' and ``ladder'' as shown
in Fig. \ref{fig: pass prod UC}. First, we select one of the microbaths,
e.g. the cold one, to set the floors and plot the level $\beta_{c}E_{\lambda}^{(c)}$
with an increasing sideways shift for each level so that a staircase
shape is obtained (the stairs may be uneven). Next on each floor we
set a ladder of the hot levels $\beta_{h}E_{\nu}^{(h)}$. The floors
are not actual levels of $\mc B$, but merely a reference for the
ladders. The positioned ladders constitute the actual levels of $\mc B$. 

With the sideways shift, it is easy to read off $\mc B$ from the
plot. Without the shift only the spectrum is accessible and it not
easy to cast it to the form $\beta_{h}H_{h}\otimes I_{c}+I_{h}\otimes\beta_{c}H_{c}$
if we do not already know $\beta_{c}H_{c}$ and $\beta_{h}H_{h}$.
Furthermore, this representation is very useful for understanding
how the spectrum changes if we continuously change some of the parameters,
for example, the temperature one of the elements. Note that the ladders
may overlap in height (Fig. \ref{fig: pass prod UC}a) or be separated
from each other (Fig. \ref{fig: pass prod UC}b). As explained later,
this separation has physical implications.

It is straightforward to extend the plot to multiple microbaths in
an iterative way. For example for three microbath, start by plotting
two as before, then consider the resulting plot as ``floors'' and
use the third microbath as ladders.Let us define a new operator $\mathcal{\widetilde{B}}$
which has the same eigenstates as $\mathcal{B}$, i.e. $[\mathcal{\widetilde{B}},\mathcal{B}]=0$,
but can have different eigenvalues. To ensure that the new operator
$\mathcal{\widetilde{B}}$ stays globally passive we want that it
will have the same order of eigenvalues as $\mathcal{B}$. Clearly,
we can move individual levels up and down and as long as we are not
crossing any levels and the order of the new eigenvalues of $\tilde{\mc B}$
will still be the same as that of $\mc B$. Crossing levels will change
the order and break global passivity. Since degenerate levels of $\mc B$
have no order between themselves, we can split and move them until
they touch another level they were not originally degenerate with.
Thus the idea of passivity deformation can be stated as follows:

\noindent\fbox{\begin{minipage}[t]{1\columnwidth - 2\fboxsep - 2\fboxrule}%
\emph{Passivity deformation}

An operator $\tilde{\mc B}$ created from $\mc B=-\ln\rho_{0}^{tot}$
using the following rules
\begin{enumerate}
\item Moving levels up and down without crossing other levels.
\item Splitting levels that were degenerate in $\mc B$.
\end{enumerate}
is globally passive, i.e. it satisfies $\Delta\left\langle \tilde{\mc B}\right\rangle \ge0$
for any thermodynamic protocol $\rho_{f}^{tot}=\sum_{k}p_{k}U_{k}\rho_{0}^{tot}U_{k}{}^{\dagger}$.%
\end{minipage}}

If the dynamics does not mix all levels (i.e. there are conserved
quantities) we have the following more flexible set of rules:

\noindent\fbox{\begin{minipage}[t]{1\columnwidth - 2\fboxsep - 2\fboxrule}%
\emph{Passivity deformation under restricted dynamics}

Let the dynamics be composed of a mixture of unitaries $U'_{k}$ that
do not mix two specific levels $m$ and $l$ (levels of the whole
setup), i.e. $\text{\ensuremath{\braOket m{U_{k}}l}}=\braOket l{U_{k}}m=0$.
An operator $\tilde{\mc B}$ created from $\mc B$ using rules 1 \&
2 above as well as the following rule
\begin{enumerate}
\item [3.]\setcounter{enumi}{1}Crossing of levels $m$ and $l$ is allowed
(provided no other level are crossed in the process).
\end{enumerate}
is globally passive, i.e., it satisfies $\Delta\left\langle \tilde{\mc B}\right\rangle \ge0$
for any thermodynamic protocol generated by the restricted dynamics
$\rho_{f}^{tot}=\sum_{k}p_{k}U_{k}'\rho_{0}^{tot}U_{k}'^{\dagger}$.%
\end{minipage}}\\

The additional third rule, can be stated using a conservation law.
In the case of a conserved quantity $Q$, crossing should be avoided
only between levels that have the same value of $Q$. 

We emphasize that these deformations are by no means physical operations
we execute on the setup. They are just a technique for finding new
globally passive operators and new inequalities. The physical interpretation
of the resulting inequalities depends on the passive operators $\tilde{\mc B}$
that can be identified. The interpretation does not have to be thermodynamic
in character (e.g. involving energies of subsystems).

As shown later in Sec. \ref{sec: non-thermal} this deformation recipe
(rules 1-3) is not limited to objects that are initially in a Gibbs
state or to uncorrelated objects. Next, we study several key examples
that illustrate the power of the passivity deformation approach and
the graphical representation in particular.

\subsection{\label{sec: Addressing-the-ultra-cold}Addressing the ultra-cold
catastrophe}

\subsubsection{Ultra-cold environments and non-overlapping ladders}

As an example for a process involving a very cold environment let
us consider once again a setup composed of two microbaths that are
initially thermal with inverse temperatures $\beta_{c}$ and $\beta_{h}$.
Then, the setup is driven by a process described by a mixture of unitaries.
If the initial temperature $1/\beta_{c}$ is too low, no unitary process
(or a mixture of unitaries) can cool the cold microbath, irrespectively
of the amount of invested work. That is, a refrigerator cannot be
implemented in the given setup. As shown below, this is a rather general
result. We term such environments ``ultra-cold environments''.

As indicated in Fig. \ref{fig: pass prod UC}a. the spacings between
the floors (dashed-blue), is proportional to $\beta_{c}$. By decreasing
the initial temperature we are expanding the spacings between the
blue levels. Yet, the spacings in the red ladder are not affected
by the change in $\beta_{c}$. Next, we assume that the spectral range
of the hot microbath $\omega_{h}^{max}=max(E^{h})-min(E^{h})$ is
finite. Under this assumption, it follows that for large enough $\beta_{c}$
the ladders will not overlap with each other as shown in Fig. \ref{fig: pass prod UC}b.
The condition for non overlapping ladders is
\begin{align}
\beta_{c} & \ge\beta_{c}^{\star},\label{eq: cold env beta*}\\
\beta_{c}^{\star} & \doteq\beta_{h}\omega_{h}^{max}/\omega_{c}^{min},
\end{align}
where $\omega_{c}^{min}=min(E_{n+1}^{c}-E_{n}^{c})$ is the minimal,
\emph{nonzero} energy gap in the cold bath Hamiltonian. Physically,
this condition is met when the initial temperature of the cold microbath
is ``sufficiently cold''. Equation (\ref{eq: cold env beta*}) implies
that this scale of coldness is not an intrinsic and objective property
of the cold environment, but a property with respect to the other
systems it potentially interacts with (the hot microbath). Note, that
in this regime the hot bath \emph{can} be very cold as well, as long
as (\ref{eq: cold env beta*}) is satisfied. Next, the relation between
no-cooling and non overlapping ladders is outlined.

\subsubsection{\label{subsec:A-no-cooling-theorem}No cooling in the ultra-cold
regime}

Considering the two-microbath scenario above, the Clausius inequality
(\ref{eq: GP microbaths}) constrains the possible changes in the
energies of the hot and cold subsystems. Yet, it does not, a priory,
determines the sign of $\Delta\left\langle H_{c}\right\rangle $ which
may depend on the selected unitary, i.e. whether an engine or a refrigerator
has been implemented.

Next, we assume that the cold environment is ultra cold and that the
no-overlap condition (\ref{eq: cold env beta*}) holds. Since the
ladders do not overlap, according to the passivity deformation rules
above, we can freely expand the distances between them (no levels
will be crossed). We uniformly increase the distance between them
by replacing $\beta_{c}$ by some fictitious $\beta_{c}'\ge\beta_{c}$.
Since this a legitimate deformation it holds that 
\begin{equation}
\beta_{c}'\Delta\left\langle H_{c}\right\rangle +\beta_{h}\Delta\left\langle H_{h}\right\rangle \ge0.\label{eq: beta_c_p no cooling}
\end{equation}
We remind the reader that this is not a physical change in the system
and the initial temperature is still $\beta_{c}$ and not $\beta_{c}'$.
Yet, we obtained a new inequality (\ref{eq: beta_c_p no cooling})
by doing this deformation. Taking the limit $\beta_{c}'\to\infty$
we conclude that
\begin{align}
\Delta\left\langle H_{c}\right\rangle  & \ge0,\\
\text{for } & \beta_{c}>\beta_{c}^{\star},
\end{align}
i.e. there is no refrigerator that can exploit the given hot environment,
to cool the given ultra-cold environment. Such behavior is known for
specific Otto engines coupled to Markovian environments \citep{k122,mahler07b,RUswap}.
Here we have used passivity deformation to show that this is a generic
property of microbaths and not of a specific machine. Even in complicated
machines that involve quantum non-adiabatic couplings that are too
complicated to be solved analytically, the conclusion on the lack
of cooling window for $\beta_{c}\ge\beta_{c}^{\star}$ still holds.
Note that there was no restriction on the applied unitary so it is
possible to add a local unitary on the cold microbath at the end of
the evolution that brings the cold microbath to its passive state.
Thus, even residual coherence cannot be utilized to overcome the $\Delta\left\langle H_{c}\right\rangle \ge0$
result for $\beta_{c}\ge\beta_{c}^{\star}$.

\subsubsection{Resolution of the zero temperature catastrophe of the second law}

As explained above when $\beta_{c}\to\infty$ the hot (finite) ladders
are infinitely separated from each other (see Fig. \ref{fig: pass prod UC}b).
According to the rules of passivity deformation described above, we
now deform the operator $\mc B$ into a new operator $\mathcal{\widetilde{B}}$
by shrinking the distance between the ladders (but not shrinking the
ladders themselves). As result, the new operator now has a fictitious
$\beta_{c}'<\infty$. According to the rules of passivity deformation
$\beta_{c}'$ cannot be arbitrary small. We must stop at the first
time the ladders cross. This happens at 
\begin{align}
\beta_{c}' & =\beta_{c}^{\star}=\beta_{h}\omega_{h}^{max}/\omega_{c}^{min}.
\end{align}
Thus, it holds that in this setup
\begin{equation}
\beta_{c}^{\star}\Delta\left\langle H_{c}\right\rangle +\beta_{h}\Delta\left\langle H_{h}\right\rangle \ge0,\label{eq: beta_c_min}
\end{equation}
or alternatively: 
\begin{equation}
\frac{1}{\omega_{c}^{min}}\Delta\left\langle H_{c}\right\rangle +\frac{1}{\omega_{h}^{max}}\Delta\left\langle H_{h}\right\rangle \ge0.\label{eq: CI no overlap}
\end{equation}
This form appears to be temperature independent, however, it is valid
only if the real initial cold temperature of the bath $1/\beta_{c}$
is smaller than $\omega_{c}^{min}/(\beta_{h}\omega_{h}^{max})$. Comparing
(\ref{eq: beta_c_min}) or (\ref{eq: CI no overlap}) to the uninformative
prediction of the second law $\Delta\left\langle H_{c}\right\rangle \ge0$
we now have a meaningful relation between the energy changes in the
cold microbath and the hot microbath. In particular, according to
(\ref{eq: CI no overlap}) the efficiency of an engine exploiting
these two environments is limited by $\eta\le1-\frac{\beta_{h}}{\beta_{c}^{\star}}=1-\omega_{c}^{min}/\omega_{h}^{max}$
while the prediction of the second law (\ref{eq: GP microbaths})
for $\beta_{c}\rightarrow\infty$ is $\eta\le1-\frac{\beta_{h}}{\beta_{c}}=1$
which provides no new useful information. The efficiency $1-\frac{\beta_{h}}{\beta_{c}^{\star}}$
corresponds to an Otto engine operating between the two levels with
the smallest nonzero energy gap in the cold bath and the two most
separated levels in the hot bath. This process will saturate the revised
bound (\ref{eq: beta_c_min}). 

We should clarify that in this scenario there is no machine that runs
a periodic protocol and achieves some steady-state operation. Instead
there is a direct non energy-conserving interaction between the two
microbaths. Yet this interaction may cool (and consume work) or harvest
some work (an engine). In the current setup there is no subsystem
that acts as a working medium that can store part of the energy. As
a result, the efficiency remains $W/Q_{h}$ even if a single shot
non-periodic drive is applied. Finally, for the readers who are familiar
with microscopic engines, we point out that small environments cannot
support the isotherm needed for the Carnot machines (see the discussion
about reversible processes in Sec. \ref{subsec: sat-path-indep}).
Thus, for small environments, the Carnot machine cannot be implemented
regardless of how slow they operate.

\subsection{\label{sec: Information-erasure}Information erasure and the thermodynamic
cost of polarization creation }

One of the advantages of the passivity deformation approach is that
it allows finding bounds on various 'fine-grained' observables and
not only on the average energy and the entropy. As an example consider
a setup composed of hot and cold microbaths. Assume that the hot microbath
is an $N$-level system where levels $m$ and $m+1$ are degenerate
$E_{m}^{h}=E_{m+1}^{h}$. The task at hand is to increase the polarization
of these two levels, i.e. to increase the population difference $\left|p_{m}^{h}-p_{m+1}^{h}\right|$
by interacting with another microbath. Note that, due to interactions
of other levels with the external driving and the other microbath,
the total average energy of the hot bath and its entropy can either
grow or decrease as the polarization is increased. Consequently, this
task is ``fined grained'', as the quantities appearing in the CI
do not set explicit bounds on the execution of such task. 

Initially, the hot microbath is in a Gibbs state and both levels are
equally populated. In the case the hot microbath is a two-level system,
polarization creation is very similar to the Landauer principle that
assign a minimal heat cost to entropy changes $\Delta S_{h}=-\beta_{c}\Delta\left\langle H_{c}\right\rangle $.
The most familiar case is a full resetting of a fully unknown state
where $S_{0}=\ln2$. Using macroscopic baths, erasure can be carried
out reversibly by protocols that combine isotherms and adiabats (both
of which satisfy zero entropy production). 

The Landauer erasure principle is one of the central results regarding
the thermodynamic consequences of handling information. It highlights
the costs that must accompany logically irreversible operations on
a subsystem. The Landauer principle is quite useful in understanding
thermodynamic processes like Maxwell demon and Szilard engine (also
known as an``information engine''). Thus, it is of interest to study
the thermodynamic cost of polarization creation which is the expectation
value analog of the Shannon/von Neumann information erasure.

\begin{figure}
\includegraphics[width=8.6cm]{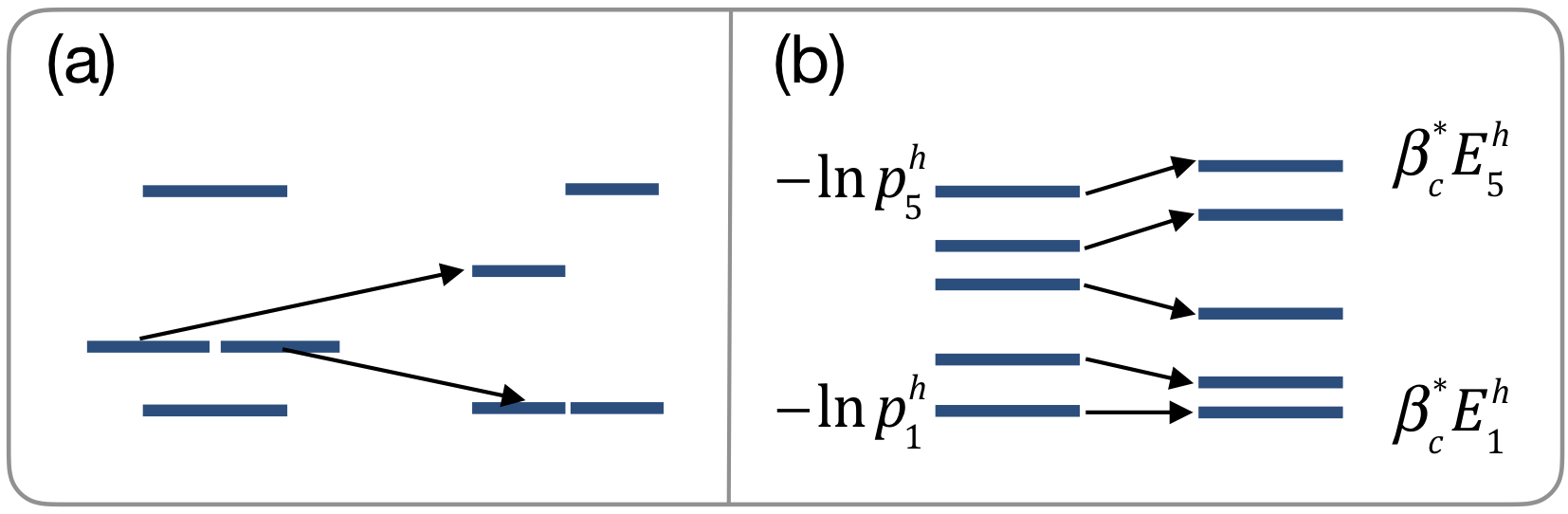}

\caption{\label{fig: erasure and athermal}(a) The deformation for obtaining
a bound on polarization erasure, in the non-overlapping ladders regime
(here only one ladder is shown). (b) The deformation that assigns
a temperature to a non thermal passive environment (one ladder is
shown - non overlapping ladders regime).}

\end{figure}

For simplicity, let us first assume that $\beta_{c}\geq\beta_{c}^{\star}$
and relax this assumption later. Moreover, for brevity, we also assume
that only two levels are degenerate ($E_{m}^{h}=E_{m+1}^{h}=E$),
and define $E_{+}\equiv E_{m+2}^{h}>E,E_{-}\equiv E_{m-1}^{h}<E$
i.e. the first upper and lower levels around the degeneracy. The population
difference of interest can be recast as the expectation value of the
operator $A=\ketbra{m+1}{m+1}-\ketbra mm$, namely $tr(\rho A)=p_{m+1}^{h}-p_{m}^{h}$.
Next, we construct the operator $\widetilde{B}(\nu)=\beta_{c}H_{c}+\beta_{h}H_{h}+\nu\beta_{h}A$,
by adding a term proportional to $A$ to the globally passive operator
$\mc B$. As shown in Fig. \ref{fig: erasure and athermal}a this
operator splits the degeneracy. 

From the no crossing rule of the passivity deformation framework we
get that $\widetilde{B}(\nu)$ is globally passive for $\nu_{-}\le\nu\le\nu_{+}$
where $\nu_{\pm}=\pm\min(E_{+}-E,E-E_{-})$. By combining the $\nu_{-}$
bound for positive changes in $A$ with the $\nu_{+}$ for negative
changes we get
\begin{equation}
\beta_{c}\Delta\left\langle H_{c}\right\rangle +\beta_{h}\Delta\left\langle H_{h}\right\rangle \geq\nu_{+}\beta_{h}\left|p_{m+1}^{h}-p_{m}^{h}\right|\label{eq: pol_ineq}
\end{equation}
where the polarization $p_{m+1}^{h}-p_{m}^{h}$ is calculated at the
end of the process (initially there is no polarization). A more restrictive
inequality can be obtained by replacing $\beta_{c}$ with $\beta_{c}^{\star}$,
resulting in
\begin{equation}
\frac{\omega_{h}^{max}}{\omega_{c}^{min}}\Delta\left\langle H_{c}\right\rangle +\Delta\left\langle H_{h}\right\rangle \geq\nu_{+}\left|p_{m+1}^{h}-p_{m}^{h}\right|.\label{eq: starred_pol_ineq}
\end{equation}

When $\beta_{c}<\beta_{c}^{\star}$ a similar inequality to (\ref{eq: pol_ineq})
holds. The only difference is in the expression for $E_{\ensuremath{\pm}}$.
For $\beta_{c}>\beta_{c}^{\star}$ it is clear that the levels in
$\widetilde{B}(\nu)$ which are the closest to the degenerate levels
$E_{\ensuremath{m}}^{h},E_{m+1}^{h}$ in a specific ladder, are in
the same ladder, since the ladders are well separated. When the ladders
overlap, the closest levels may originate from different ladders.
Nonetheless, the principle remains the same, and $\nu_{\pm}$ is obtained
from the maximal degeneracy splitting before the nearest level is
crossed. Note that for $\beta_{c}<\beta_{c}^{\star}$ it is not possible
to replace $\beta_{c}\to\beta_{c}^{\star}$ as done in (\ref{eq: starred_pol_ineq})
for the case where the ladders do not overlap.

We stress that the bound (\ref{eq: starred_pol_ineq}) is tight for
small environments, in contrast to the Landauer bound(see appendix
III for the reason the second law and the Landauer bound are not tight
when the environment is microscopic). Furthermore, if the dynamics
respects some conservation laws that exclude the possibility of executing
the bound saturating operation (Sec. \ref{subsec: sat-path-indep})
we can incorporate the conservation law into (\ref{eq: pol_ineq})
or (\ref{eq: starred_pol_ineq}) as done in Sec. \ref{sec: PD-graphic}
(second box). As a result, a new attainable bound is obtained where
the value of the new $\nu_{+}$ is greater than the value of $\nu_{+}$
without the conservation law.

\subsection{\label{sec: non-thermal}Local and global deviations from initial
local equilibrium}

In the previous sections, the goal was to find new inequalities for
initial states of several uncoupled and thermal subsystems, where
the Clausius inequality holds. We now do the opposite and look for
scenarios where the initial conditions are such that the second law
(\ref{eq: GP microbaths}) may be violated (e.g. initially correlated
microbath). Our goal is to use passivity deformation to obtain bounds
that are valid in this regime, yet still have the same form of the
second law, given by(\ref{eq: GP microbaths}), albeit involving some
effective temperatures.

\subsubsection{Initially athermal passive subsystem}

Consider a situation in which a subsystem is prepared in state $\rho_{0}^{pass}$
which is athermal, yet still passive with respect to its Hamiltonian
$H_{s}$, .i.e. $p_{n}\le p_{k}$ if $E_{n}^{s}\geq E_{k}^{s}$ and
$[-\ln\rho_{0}^{pass},H_{s}]=0$. This athermal system is then coupled
to an ultra cold microbath by a process that is described by a mixture
of unitaries. The global passivity of $-\ln\rho_{0}^{tot}$ means
that $\Delta\left\langle -\ln\rho_{0}^{pass}\right\rangle +\beta_{c}\Delta\left\langle H_{c}\right\rangle \ge0$.
Unfortunately, this expression provides a prediction on the observable
$-\ln\rho_{0}$ and not on the energy of the initially passive system.
Is it possible to derive an inequality that would constrain the variation
of this energy?

Since the ladders are now given by the expression $-\ln\rho_{0}^{pass}$
instead of $\beta_{h}H_{h}$, the non-overlapping ladders condition
now reads
\begin{equation}
\beta_{c}\ge\bar{\beta}_{c}^{\star}\doteq\frac{1}{\omega_{c}^{min}}\ln p_{1}/p_{N}.
\end{equation}
In what follows we assume that this condition holds. Consequently,
it is possible to use the passivity deformation rules and get a new
globally passive operator using the deformation $-\ln p_{i}\to\beta_{s}^{eff}H_{s}$
as depicted in Fig. \ref{fig: erasure and athermal}b. The value of
$\beta_{s}^{eff}$ is determined by the no overlap condition $\beta_{s}^{eff}\omega_{s}^{max}=\beta_{c}\omega_{c}^{min}$.
The resulting bound is
\begin{equation}
\frac{1}{\omega_{c}^{min}}\Delta\left\langle H_{c}\right\rangle +\frac{1}{\omega_{s}^{max}}\Delta\left\langle H_{s}\right\rangle \ge0.\label{eq: Athermal CI}
\end{equation}
We have demonstrated that such athermal initial states still leads
to bounds that restrict energy exchanges. For example, in the no overlap
regime an engine exploiting this passive environment is limited by
the Otto efficiency with a compression ratio of $\omega_{c}^{min}/\omega_{s}^{max}$. 

The extension to two athermal small environments in the non-overlapping
ladders regime is straightforward. Denoting the additional athermal
environment that replaces the cold microbath by s', one gets that
(\ref{eq: GP microbaths}) remains valid (replacing labels '$c$'
by '$s'$') valid, but now the no-overlap condition reads
\begin{equation}
\min(\ln p_{n}^{s'}/p_{n-1}^{s'})\ge\ln p_{1}/p_{N}
\end{equation}
Since (\ref{eq: Athermal CI}) holds in this case as well, the Otto
efficiency limits the performance of the engine even though non of
the environment is initially thermal. Moreover, exactly as in Sec.
\ref{subsec:A-no-cooling-theorem}, since the ladders do not overlap
it is not possible to cool and reduce the average energy of $s'$.

\subsubsection{``Inert'' classical correlations between subsystems}

Next, we consider a scenario in which there are initial correlations
(in the energy basis) between different subsystems. For concreteness,
we consider a setup with the initial Hamiltonian $H=H_{c}+H_{h}$.
The initial state of the setup is diagonal in the eigenbasis of the
Hamiltonian, but does not form a product state, namely $p_{ij}^{\text{ch}}\neq p_{i}^{c}p_{j}^{h}$
where $p$ are the diagonal elements of the density matrix. As before,
the setup is driven by a process  modeled by a mixture of unitaries.
At the end of the process, the coupling is turned off so that the
final Hamiltonian is the same as the initial one.

The initial correlations of the type considered here can be achieved
by creating interaction between subsystems and turning it off before
the process starts. In quantum setups, such a procedure may result
in additional quantum correlations. Nevertheless, there are situations
where the quantum correlation reduces to classical correlation. For
example, if two objects with different energy gaps are brought momentarily
into resonance by a driving field, the free evolution after the drive
is switched off, will cause the off-diagonal elements of the density
matrix to rotate in time. If it is not known when the correlating
interaction took place, only the time-averaged density matrix is accessible.
As a result, the time-averaged state is classically correlated. Alternatively,
if the two objects are subjected to slow local dephasing after the
correlating interaction is switched off, the joint state will be classically
correlated.

\begin{figure}
\includegraphics[width=8.6cm]{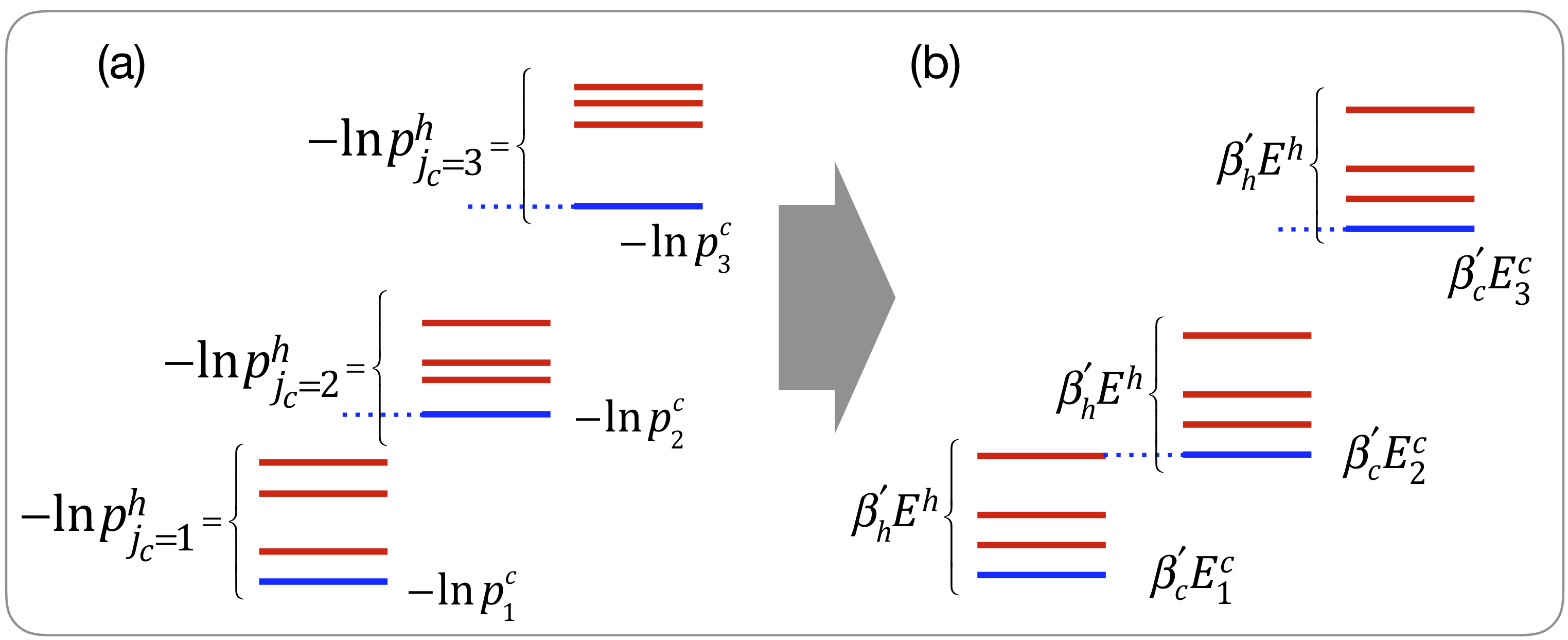}

\caption{\label{fig: corr IC-1}(a) When two environments are initially classically
correlated, the globally passive operator $\protect\mc B=-\ln\rho_{0}^{pass}$
no longer has the 'ladder replicas' structure shown in Fig\ref{fig: pass prod UC}.
Instead, each ladder is determined by the conditional probability
via $-\ln p(E_{h}|E_{c})$. (b) If the manifolds are separated as
shown in (a), then it is possible to deform the original $\protect\mc B$
into a new operator with a replica structure and thermal local operators
that yield a second law of the form $\beta_{c}^{eff}\Delta\left\langle H_{c}\right\rangle +\beta_{h}^{eff}\Delta\left\langle H_{h}\right\rangle \ge0$.
This procedure yields an effective temperature inspired by the initial
classical correlation.}
\end{figure}

In the derivation of Eq. (\ref{eq: GP microbaths}) the lack of correlations
is a crucial assumption (see \citep{MySecondLawReview} and reference
therein). Indeed such initial correlations may result in energy flow
between subsystems \citep{Jennings2010ReverseFlow} that contradict
the second law. Below, we show that if one subsystem is sufficiently
cold a second-law like inequality holds, but with effective temperatures
that depend on the initial correlations. We start by writing the classically
correlated initial state as
\begin{equation}
p_{ij}^{ch}=p_{i}^{c}p_{j|i}^{h|c}\label{eq: conditional form}
\end{equation}
where $p_{j|i}^{h|c}$ is the conditional distribution of the hot
environment given the state of the cold environment. By taking $-\ln$
of the right hand side of (\ref{eq: conditional form}) it becomes
clear how to extend the ``floors and ladders'' diagram to the classically
correlated case: $\{-\ln(p_{i}^{c})\}_{i}$ constitute the floors,
and they are used for plotting the initial staircase as before. Next,
on floor $i$ we put the ``conditional ladder'' $\{-\ln(p_{j|i}^{h|c})\}_{j}$
that corresponds to it. That is, the correlation manifest in the fact
that the ladders are different from each other.

In this way, arbitrary strong classical correlation can be represented.
In order to apply passivity deformation and derive a bound that resembles
(\ref{eq: GP microbaths}), we make the following assumptions: 1)
As in the previous section, the ladders do not overlap; 2) The conditional
marginals $\{-\ln(p_{j|i}^{h|c})\}_{j}$ are passive with respect
to $H_{h}$, and 3) $\{-\ln(p_{i}^{c})\}_{i}$ is passive with respect
to $H_{c}$. We do not assume the marginals are thermal or that the
correlation is weak.

Starting with the cold subsystem, we first shift the floors from $\{-\ln(p_{i}^{c})\}_{i}$
to $\beta_{c}^{eff}H_{c}$. Since the ladders do not overlap there
is always some large enough $\beta_{c}^{eff}$ for which this is possible.
To obtain a tight bound, we select the minimal value for which this
is possible. Next, to get rid of the correlation in our constructed
globally passive operator, we need to make all the ladders identical.
Since we assumed that the conditioned marginals of the hot object
are passive, it implies that each ladder has the same ordering as
$\beta_{h}H_{h}$. Therefore, we can use the ladder separation to
deform each ladder separately into $\beta_{h}^{eff}H_{h}$. For achieving
the best bound $\beta_{h}^{eff}$ is taken to be the largest value
that does not make the ladders cross. 

We conclude that if the three conditions above hold, then despite
the initial correlations a second-law-like bound of the form 
\begin{equation}
\beta_{c}^{eff}\Delta\left\langle H_{c}\right\rangle +\beta_{h}^{eff}\Delta\left\langle H_{h}\right\rangle \ge0,
\end{equation}
holds for any mixture of unitaries. The information about the correlations
is encapsulated in the values of the effective temperatures. Note,
that by construction this bound is tight (see Sec \ref{subsec: sat-path-indep}). 

As in Sec. \ref{subsec:A-no-cooling-theorem}, the no-overlap condition
leads to $\Delta\left\langle H_{c}\right\rangle \ge0$. Hence, we
call such a correlation ``inert correlation'' as it cannot be exploited
to cool the cold environment. If, however, the product state created
from the product of the marginals $\rho_{prod}=tr_{h}\rho_{ch}\otimes tr_{c}\rho_{ch}$
satisfies the no overlap condition but $\rho_{ch}$ does not (non
inert case), it is possible to cool the object that appears to be
locally ultra-cold.

Several recent papers were devoted to the study of fluctuation theorems
of small systems that are strongly coupled to their environment \citep{SeifertPRL2016_strong_coupling,JanetPRE2017_strong_coupling,Jarzynski2017_PRX_strong_coupling,Esposito_PRE2017_strong_coupling}
. These inevitably involve significant classical correlations between
the system and its environment. One should note several important
differences between the setups. Specifically, in the study of fluctuation
theorems, the coupling between a system and its environment is always
present, so the question of dividing the interaction energy between
subsystems becomes non trivial. In addition, the framework of stochastic
thermodynamics deals with fluctuating quantities. Our results pertain
to changes of expectation values.

\section{\label{sec: Coarse-graining}Coarse-graining, the truncated CI, and
the binary CI }

In this last part of the paper we present three bounds whose utility
stems from intentionally ignoring the distinguishability of some energy
levels. The first bound deals with situations in which one wishes
to ignore certain degrees of freedom and corse-grain the setup. The
other two inequalities are relevant for heat leak detection and lazy
demon detection. Moreover these last two inequalities reveal a hierarchical
structure that starts with the second law and ends with a majorization
condition.

\subsection{Coarse-graining}

Consider the case where the energy levels in the setup of interest
have some internal structure that our detectors cannot resolve. We
describe this physical situation by considering a model with two sets
of quantum numbers, so that the energy eigenstates can be denoted
by $\ket{n,m}$. Here $n$ corresponds to the resolvable degrees of
freedom, while $m$ refers to the experimentally unresolvable internal
structure. Our goal is to write a thermodynamic inequality that depends
only on the measurable degrees of freedom $n$.

Let $\rho_{0}^{full}=\sum_{n=1}^{N}\sum_{m}^{M_{n}}p_{nm}\ket{n,m}\bra{n,m}$
be the full initial density matrix of the setup. Global passivity
leads to $\Delta\left\langle \mc B^{full}\right\rangle \ge0$ where
$\mc B^{full}=-\ln\rho_{0}^{full}$. This inequality clearly depends
on all the degrees of freedom of the setup including the internal
ones that we wish to coarse grain. We ask under what conditions it
holds that 
\begin{equation}
\Delta\left\langle \mc B^{CG}\right\rangle \ge0\label{eq:CGinequality}
\end{equation}
where $\mc B^{CG}=\sum_{n=1}^{N}q_{n}\ket n\bra n$, and $q_{n}$
are some real numbers that we will obtain shortly. 

It is useful to start by examining a special case where the coarse-graining
is straight forward. Assume that all the internal levels are degenerate
i.e. $E_{n,m}=E_{n}$. This implies that the initial probabilities
are only function of $n$, $p_{nm}^{0}=\widetilde{p}_{n}$ (the dimension
of $\widetilde{p}_{n}$ is $N$). This degeneracy enables to simplify
of the inequality $\Delta\left\langle \mc B^{full}\right\rangle \ge0$,

\begin{align}
0 & \leq\triangle\left\langle \mc B^{full}\right\rangle =\sum_{nm}\left(p_{nm}^{f}-p_{nm}^{0}\right)\left[-\ln p_{nm}^{0}\right]\nonumber \\
 & =\sum_{n}\left(p_{n}^{f}-p_{n}^{0}\right)\left[-\ln\widetilde{p}_{n}\right]=\triangle\left\langle \mathcal{B}^{CG}\right\rangle ,\label{eq: CGderive}
\end{align}
where $p_{n}^{f,0}\equiv\sum_{m}p_{nm}^{f,0}$ and $\mathcal{B}^{CG}=-\sum\ln\widetilde{p}_{n}\ket n\bra n$.

\begin{figure}
\includegraphics[width=8.6cm]{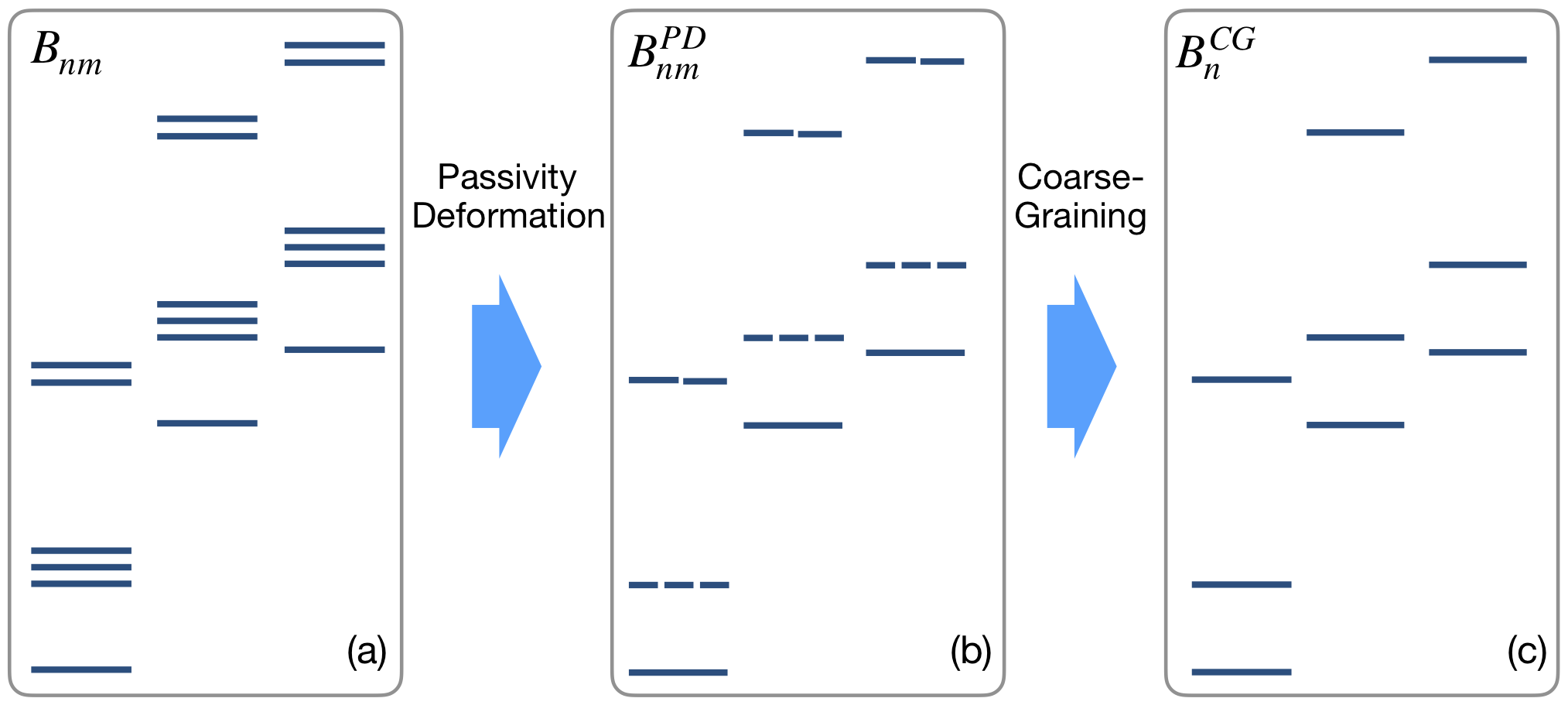}

\caption{\label{fig: coarse grain}(a-b) the passivity deformation that makes
all the immeasurable degrees of freedom degenerate. (b-c) now it is
possible to coarse grain and ignore the internal structure of the
levels. This process is possible only when the internal manifolds
do not overlap. This is a weaker condition compared to the full no-overlap
situation shown in Fig. \ref{fig: pass prod UC}b.}

\end{figure}

When the probabilities $p_{nm}^{0}$ are not degenerate in the quantum
number $m$, the passivity deformation approach can be used to find
out if, and under what conditions one can obtain a coarse-grained
inequality. Consider the case shown in Fig.\ref{fig: coarse grain}a
where the values $\mathcal{B}_{nm}=-\ln p_{nm}^{0}$ are clustered,
with different values of $n$ denoting different clusters, while $m$
differentiate between states in the same cluster. Passivity deformation
allows deform $\mathcal{B}_{nm}$ into a new passive operator $\mathcal{\widetilde{B}}_{nm}$
that is independent of $m$, as depicted in Fig.\ref{fig: coarse grain}b.
Then, one can repeat the argument in (\ref{eq: CGderive}) and obtain
a lower-dimensional operator that satisfies the effective inequality
(\ref{eq:CGinequality}). This coarse-graining procedure is allowed
\emph{as long as the different clusters do not overlap}. The result
is not unique, since the value of $q_{n}$ can be changed (as long
as their order is kept) without affecting the validity of (\ref{eq:CGinequality}).
This, however, is consistent with the low resolution of the detector
that motivated the coarse-graining to begin with.

Note that in general, the passivity-based inequality $\Delta\left\langle \mathcal{B}^{CG}\right\rangle \ge0$
is different from the one obtained by first coarse-graining the probability
distribution and then applying passivity. If the coarse-graining is
done first one finds$B_{n}'=-\log p_{n}^{0}=-\log\sum_{m}p_{n,m}^{0}=-\log M_{n}\widetilde{p}_{n}$
where $M_{n}$ is the degeneracy of level $n$ . If $M_{n}=M$ i.e.
all the $n$ levels have the same degeneracy, then $B_{n}'=-\log\widetilde{p}_{n}+\text{const }$
where the additive constant can be ignored as it drops out when calculating
$\Delta\left\langle B'\right\rangle $. We conclude that if $M_{n}=M$,
it holds that $\mc B_{n}^{CG}=B'+\text{const. }$ That is in this
special case it it not important if the coarse-graining is done in
probability space or in the passive operator space. Yet, generally,
$\mc B_{n}^{CG}$should be used and not $B'$. For example, for a
thermal state with some degenerate energy structure $E_{nm}=E_{n}$
we get $\mc B_{n}^{CG}=\beta E_{n}$ while $B'_{n}=\beta E_{n}-\log M_{n}$.
However, passivity provides an inequality involving $\mc B_{n}^{CG}$
and not $B'_{n}$. Thus, this example illustrates the importance of
coarse-graining the passive operator and not the probabilities.

\subsection{The truncated second law}

In the following, we show a generic prescription (i.e. one that can
be carried out in any setup) for constructing inequalities that depend
only on parts of the Hilbert space, while ignoring others. These inequalities
can exhibit superior heat leak detection, or lazy demon detection. 

Let us consider a setup with several initial uncorrelated microbaths.
We start by considering the operator $\mathcal{B}=-\ln\rho_{0},$
that is globally passive by construction. Denoting by \textbf{$0\le b_{1}\le b_{2}\le b_{3}..$}
the sorted eigenvalues of $\mc B$ (``levels'' of $\mc B$), let
us apply the following deformation: first, the lowest level $b_{1}$
is lowered to zero. Since the eigenvalues of $b_{i}\ge0$, this means
to move the level to a point which is lower than all the other levels.
Then, the second-lowest level $b_{2}$ is lowered to zero, and this
is repeated until only the $l$ highest levels (i.e. the least-populated
levels) remain. According to passivity deformation, the resulting
operator $\mc B^{(l)}=\sum_{k=N-l+1}^{N}b_{i}\ketbra{b_{i}}{b_{i}}$
satisfies 
\begin{equation}
\Delta\left\langle \mc B^{(l)}\right\rangle \ge0.\label{eq: partial B ineq}
\end{equation}
The physical interpretation of such an observable is not self-evident.
For example, initially uncorrelated microbath $\mc B$ can be written
as a sum of local operators $A_{c}\otimes I_{h}+I_{c}\otimes A_{h}$,
but $\mc B^{(l)}$ in general cannot be written in this form since
the remaining ``ladders'' are not identical. Nevertheless, for a
setup composed of several microbaths $\mc B^{(l)}$ are observables
in the energy basis, and as such, they can be obtained from energy
measurements in the different subsystems.

We have numerically verified that in a linear spin chain where a lazy
demon operates between the two middle spins, the $\mc B^{(l)}$ inequalities
can detect demons that $\text{sign}(\alpha)B^{\alpha}$ cannot detect.
This was checked for chains of up to ten spins. At first, it may seem
surprising that by cropping a few levels from $\mc B$ the sensitivity
may be improved. However, while some levels of $\mc B$ are more affected
by feedback or heat leaks, some are not affected, and contribute a
positive value to $\Delta\left\langle \mc B\right\rangle \ge0$. This
positive contribution can degrade the detection. By excluding these
levels (within the limitations of the passivity deformation rules)
better sensitivity may be achieved (depending on the specific demon
mechanism). 
\begin{figure}
\includegraphics[width=8.6cm]{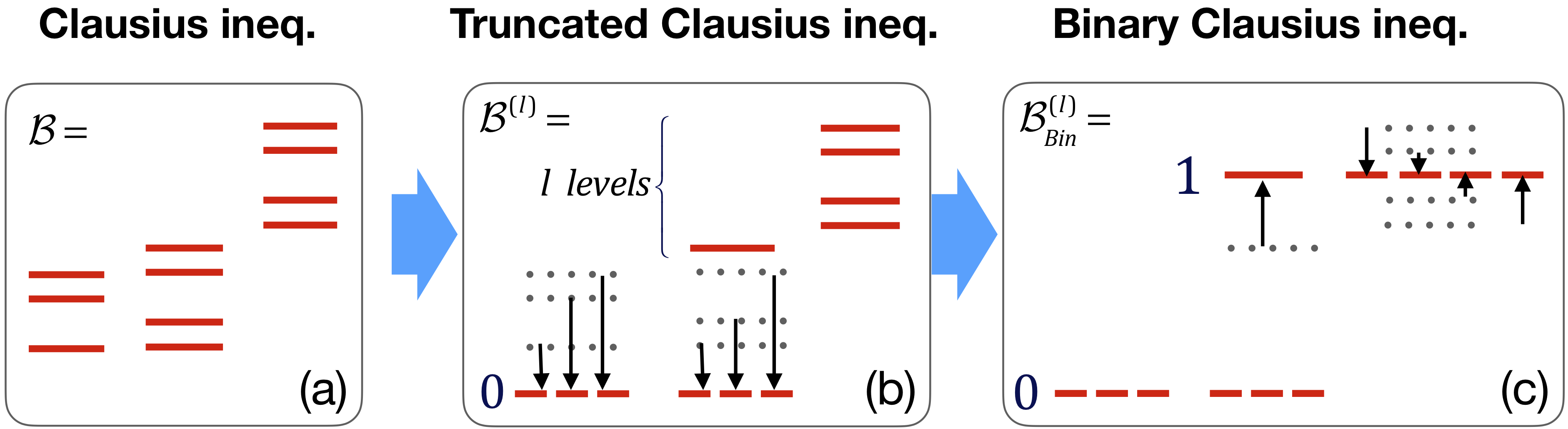}\caption{(a-b) the deformation that leads to the truncated Clausius inequality.
(b-c) The deformation that leads to binary Clausius inequality.}

\end{figure}

\subsection{Binary inequalities and their relation to majorization}

Next, we study a similar deformation and relate it to majorization.
By applying the same logic as in the previous deformation, the lowest
$N-l+1$ level can be shifted to zero as before and the highest $l$
levels to the value of one. The resulting passivity deformation inequalities
are
\begin{equation}
\Delta\left\langle \mc B_{Bin}^{(l)}\right\rangle \ge0,\label{eq: pseudo majorization}
\end{equation}
where $\mc B_{Bin}^{(l)}=\sum_{k=N-l+1}^{N}\ketbra{b_{i}}{b_{i}}$
are projection operators. These observables are binary. They only
test if our system is in a certain subspace of the Hilbert space.
In practice, energy is measured in all microbaths - if the value is
in the space of $\tilde{\mc B}_{l}$, then $\tilde{\mc B}_{l}$ take
the value of one, otherwise it is zero. At first, these binary inequalities
look even stranger than (\ref{eq: partial B ineq}), but they can
be understood directly from the following majorization relation between
the initial and final populations in the basis of $\mc B$. By construction,
the initial density matrix is diagonal in the basis of $\mc B$. Thus,
for any evolution (\ref{eq: mix uni 2}) the final populations in
this basis $\{p_{i}^{f}\}$ are related to initial ones $\{p_{i}^{0}\}$
(i.e. the eigenvalues) through the majorization relation (Schur lemma
\citep{Marshall1979MajorizationBook})
\begin{equation}
\sum_{j=1}^{l}p_{\uparrow,j}^{0}\le\sum_{j=1}^{l}p_{\uparrow,j}^{f}\:\:\:\text{for any }1\le l\le N,\label{eq: pop maj}
\end{equation}
where the up arrow stands for increasing order. First, we notice that
\begin{equation}
\sum_{j=1}^{l}p_{\uparrow,j}^{0}=tr[\rho_{0}^{tot}\mc B_{Bin}^{(l)}]\triangleq\left\langle \mc B_{Bin}^{(l)}\right\rangle _{0},\label{eq: p0 B0}
\end{equation}
and also that 
\begin{equation}
\sum_{j=1}^{l}p_{\uparrow,j}^{f}\le\sum_{j=1}^{l}p_{j}^{f}=\left\langle \mc B_{Bin}^{(l)}\right\rangle _{f}\label{eq: pf}
\end{equation}
Consequently, from (\ref{eq: pop maj})-(\ref{eq: pf}) it follows
that

\begin{align}
\left\langle \mc B_{Bin}^{(l)}\right\rangle _{0} & =\sum_{j=1}^{l}p_{\uparrow,j}^{0}\le\sum_{j=1}^{l}p_{\uparrow,j}^{f}\nonumber \\
 & \le\left\langle \mc B_{Bin}^{(l)}\right\rangle _{f}\:\:\:\text{for any }1\le l\le N
\end{align}
which yields (\ref{eq: pseudo majorization}). Conditions (\ref{eq: pseudo majorization})
are identical to majorization (\ref{eq: pop maj}) when the final
populations have the same ordering as the initial population (and
then $\sum_{j=1}^{l}p_{\uparrow,j}^{f}=\left\langle \mc B_{Bin}^{(l)}\right\rangle _{f}$).

Since any $\Delta\left\langle \mc B^{(l)}\right\rangle \ge0$ (\ref{eq: partial B ineq})
can be written as a convex sum of several $\Delta\left\langle \mc B_{Bin}^{(l)}\right\rangle \ge0$
inequalities, it follows that any violation of (\ref{eq: partial B ineq})
is associated with a violation of at least one of the inequalities
(\ref{eq: pseudo majorization}), but not necessarily the other way
around. The family of inequalities (\ref{eq: pseudo majorization})
is more sensitive than (\ref{eq: partial B ineq}). That being said
(\ref{eq: pseudo majorization}) are void of any energetic meaning.
In particular, it does not reduce to the standard second law under
some assumptions, e.g. setting $l=N$ as in (\ref{eq: partial B ineq}).

Finally, we point out that the CI, truncated CI, binary CI and Majorization
(in this order) form a hierarchal structure that reflects a tradeoff
between physical context (maximal for the CI) and tightness (maximal
for the Majorization).

\section{Conclusion}

In this paper, we have presented a framework that enables to derive
thermodynamic-like inequalities to fine-grained observables that these
days become measurable in various setups such as ion traps, optical
lattices, superconducting qubits and more. These observable do not
appear in the second law and are therefore not constrained by it.
On top of being applicable to new observables, the passivity deformation
framework also overcomes some of the deficiencies of the second law:
it yields tight bounds in scenarios where the second law is not tight,
it can be integrated with conservation laws, and it produces meaningful
results even when one of the environments is very cold.

What makes all this possible is the exploitation of the energy spectrum
of the various elements in the setup. While the use of such information
makes no sense in macroscopic systems, in microscopic systems it is
quite natural as the Hamiltonians are typically known (especially
in man-made setups).

In the future, heat machines may not be restricted to cooling and
work extraction. For example, quantum machines can be used to build
up entanglement or to manipulate some observables that are not directly
associated with energy or entropy. The upper bounds on the performance
of such machines calls for new thermodynamic theories. In this work,
we have used passivity deformation to set bounds on the performance
of such machines.

We have illustrated that passivity deformation does not only produce
bounds but also insights. By exploring the tightness of the bounds,
path-independent processes with positive entropy production were identified.
In addition, we identified scenarios where processes with athermal
and correlated environments satisfy the standard second law with some
effective temperature. Moreover, passivity deformation offers a clear
recipe for coarse-graining and provides the conditions for the validity
of bounds when some degrees of freedom cannot be resolved in the measurement
process.

In the next stages of the theory, it would be interesting to explore
the application of the theory to setups with particle transport, markovian
limits, or steady-state operation. It is also of interest to understand
the scaling behavior as the microbaths size increases. It is also
of interest to investigate to what extent these inequalities can reveal
that a system is not well isolated from the rest of the world \citep{Deffner2018ErrorDetectionAnnealers},
e.g. due to the presence of heat leaks or lazy Maxwell demons \citep{GlobalPassivity}.
The detection of isolation breach can be used to compare the predictive
power of different thermodynamic theories such as passivity deformation,
thermodynamic resource theory, and stochastic thermodynamics \citep{uzdin2019IBMexp}.
Moreover, passivity deformation bounds can potentially be used to
check coding errors in simulations, quality of devices, consistency
of various approximations, and the validity of experimental data (e.g.,
in the case of potential forgery). Our findings can be verified in
various quantum setups such as ion traps, neutral atoms in optical
lattices, or in presently available superconducting quantum processors
\citep{IBMintroWeb}. A proof of principle experimental demonstration
of superior heat detection based on passivity deformation was successfully
carried in out in a companion paper using the IBM quantum processors
\citep{uzdin2019IBMexp}.

\section*{ACKNOWLEDGMENTS}

SR is grateful for support from the U.S.-Israel Binational Science
Foundation (Grant No. 2014405), and from the Israel Science Foundation
(Grant No. 1526/15).

\section*{Appendix I - passivity as a binary relation}

To methodically study the global passivity of various operators with
respect to $\rho_{0}$ we introduce in this appendix the ordering
function tool. Let $A$ and $B$ ($B$ is unrelated to $\mc B$) be
two Hermitian matrices of the same dimensionality, the (mutual) ordering
function is given by
\begin{equation}
\chi(A,B)\doteq\text{tr}(AB)-\lambda_{A}^{\downarrow}\cdot\lambda_{B}^{\downarrow}.
\end{equation}
where $\lambda_{A(B)}^{\downarrow}$ are the eigenvalues of $A(B)$
sorted in a decreasing order. $A$ and $B$ have the \emph{same ordering}
if and only if
\begin{equation}
\chi(A,B)=0.\label{eq: general ordering condition}
\end{equation}
Similarly A and B have reverse ordering if and only if

\begin{equation}
\chi^{\downarrow\uparrow}(A,B)=0,
\end{equation}
where the reverse ordering function is
\begin{equation}
\chi^{\downarrow\uparrow}(A,B)\doteq\text{tr}(AB)-\lambda_{A}^{\downarrow}\cdot\lambda_{B}^{\uparrow}.
\end{equation}
In this notation, the two conditions for global passivity in (\ref{eq: pass state def B})
can be jointly written as
\begin{equation}
\chi^{\downarrow\uparrow}(A,\rho_{0}^{tot})=0.\label{eq: gp cond rev ord}
\end{equation}
The two ordering functions have some useful properties. Let $f_{\ensuremath{C}}$
be a strictly monotonic decreasing function $f_{\ensuremath{C}}'(x)<0$
in the spectral range of operator C, $x\in(min(\lambda_{C}),max(\lambda_{C}))$,
and similarly $g_{C}$ satisfies $g_{\ensuremath{C}}'(x)>0$ in the
same regime then it holds that
\begin{align}
\chi^{\downarrow\uparrow}(A,B)=0 & \iff\chi(f_{A}(A),B)=0\nonumber \\
 & \iff\chi(A,f_{B}(B))=0\label{eq: order to rev order}\\
\chi(A,B)=0 & \iff\chi(A,g_{B}(B))=0\nonumber \\
 & \iff\chi(g_{A}(A),B)=0\label{eq: monotonicity of ord}
\end{align}
Using $f_{B}(B)=-\ln B$ in (\ref{eq: order to rev order}) we conclude
that the global passivity condition (\ref{eq: gp cond rev ord}) for
operator $A$ can be written as
\begin{equation}
\chi(A,\mc B)=0\text{ (global passivity of A)}
\end{equation}
where $\mc{B=-\ln}\rho_{0}^{tot}$ as before (\ref{eq: B0 def}).
As a reassurance exercise, we set $A=\mc B$ and get that $\mc B$
is globally passive, since any operator is ordered with respect to
itself. Next we use property (\ref{eq: monotonicity of ord}) to deduce
\begin{equation}
\chi(g_{\mc B}(\mc B),\mc B)=0,
\end{equation}
from $\chi(\mc B,\mc B)=0$. Choosing $g_{\mc B}(x)=\text{sign}(\alpha)x^{\alpha}$
we get the globally passivity inequalities 
\begin{equation}
\Delta\left\langle \text{sign}(\alpha)\mc B^{\alpha}\right\rangle \ge0
\end{equation}
for any evolution of the form (\ref{eq: mix uni 2}) in the setup
\citep{GlobalPassivity}.

Ordering as defined in (\ref{eq: general ordering condition}) can
be viewed as binary relation: $A\sim B\Longleftrightarrow\chi(A,B)=0$.
While this binary relation is reflexive $A\sim B$, and symmetric
$A\sim B=B\sim A$ it may not be transitive, i.e. it is possible that
$A\sim B$ and $B\sim C$ but $A\cancel{\sim}C$. Hence, in general
$\chi=0$ is not an equivalence relation.

The breakdown of transitivity takes place if $A$ and $B$ have different
degeneracy structure. We say that $B$ is non-equivalent to $A$ if
1) $A\sim B$ and 2) at least two eigenvectors that are \emph{non
degenerate} in $A$, are degenerate in $B$, i.e. there are at least
two eigenvectors $v_{k},v_{l}$ such that 
\begin{align}
\braOket{v_{k}}A{v_{k}} & \neq\braOket{v_{l}}A{v_{l}},\\
\braOket{v_{k}}B{v_{k}} & =\braOket{v_{l}}B{v_{l}}.
\end{align}
Now, it is easy to construct and operator $C$ that satisfies $B\sim C$
but not $A\sim C$. As an example for $j\neq k,l$ we choose $\braOket{v_{j}}C{v_{j}}=\braOket{v_{j}}A{v_{j}}$
and for $k$ and $l$ we set $\braOket{v_{l}}C{v_{l}}=\braOket{v_{k}}A{v_{k}}$
and $\braOket{v_{k}}C{v_{k}}=\braOket{v_{l}}A{v_{l}}$. By construction
$A\cancel{\sim}B$ since the ordering is opposite in k and l. Yet,
due to the degeneracy in B it holds that $B\sim C$.

This example explains why in the deformation rules in Sec. \ref{sec: PD-graphic}
we can only split degeneracies that were already there in the initial
density matrix. Otherwise, it would contradict the non-crossing rule:
we can first make the two states degenerate and then split them in
the other direction which creates a forbidden crossing. 

\section*{Appendix II - From passivity to optimal protocols of various machines
and tasks}

Consider a setup that aims to achieve a maximal change in the expectation
value of a certain observable of interest $A$ (an Hermitian operator).
The observable may be 'local' i.e. involve only one element of the
setup or it may be global and involve several elements or even the
whole setup. For example, in refrigerators the goal is to minimize
the average energy of a cold subsystem$\text{\ensuremath{\left\langle A\right\rangle }}=\left\langle H_{c}\right\rangle $,
which is a local quantity. In engines, the goal is to reduce the energy
of the whole setup $\text{\ensuremath{\left\langle A\right\rangle }}=\left\langle H_{tot}\right\rangle $
(global quantity), since this change is equal to the amount work exchanged
with the driving field that executes the protocol. Note, however,
that $A$ does not have to be related to energy or to the original
basis of the initial state of the setup. $A$ can be any hermitian
operator bounded from below in the Hilbert space of the setup (e.g.
see the dephasing example in Sec. \ref{subsec: dephasing}).

Let us assume that the initial state of the setup $\rho_{0}^{tot}$
is given, and so is the operator $A$ that describes the observable
of interest. Thus, the initial expectation value of $A$, $\left\langle A\right\rangle _{0}=tr[\rho_{0}^{tot}A]$
is fixed. Our goal is to find the optimal unitary $U_{opt}$ that
will produce the lowest value of $\left\langle A\right\rangle $ i.e.
$A_{0}\to A_{min}=tr[\rho_{opt}^{tot}A]=tr[U_{opt}\rho_{0}^{tot}U_{opt}^{\dagger}A]$.
Fortunately, this problem is already solved by the principle of passivity.
Adopting the logic of passivity, finding $U_{opt}$ is simple, the
unitary that transforms $\rho_{0}^{tot}$ into a passive state with
respect to $A$ : $\chi^{\downarrow\uparrow}(\rho_{opt}^{tot},A)=0$
will do the job. This is can be carried out in two steps. The first
step is to rotate $\rho_{0}^{tot}$ to the basis of $A$ (if it is
not already in this basis). The second step is to apply simple level
permutations that will rearrange the populations in a monotonically
decreasing order with respect to the eigenvalue of $A$.

\begin{figure}
\includegraphics[width=8.6cm]{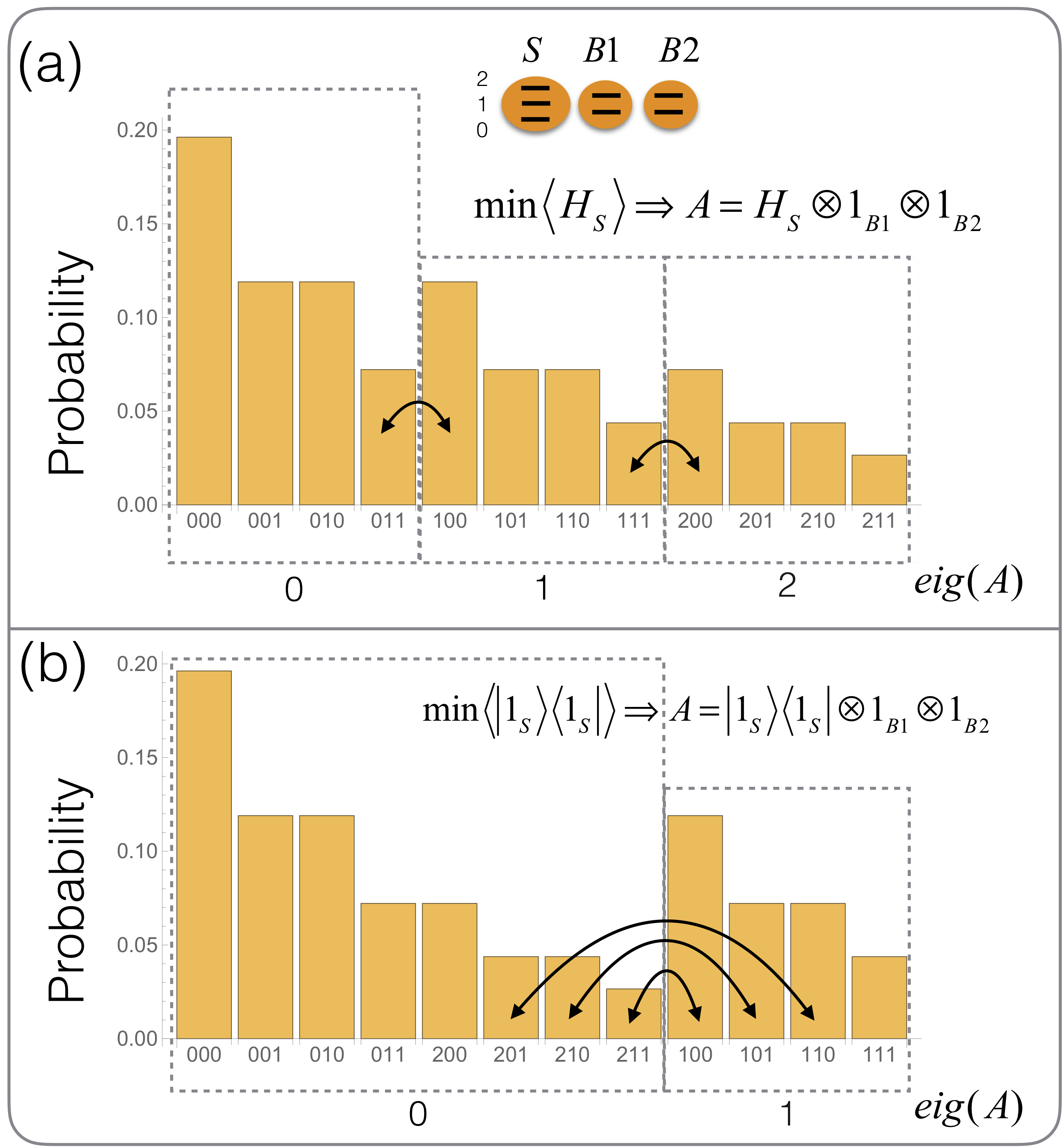}

\caption{\label{fig: Optimal-protocol}Optimal protocol for manipulating a
qutrit system (S) using two qubits (B1 \& B2). (a) To reduce the average
energy of S the probability distribution of the eigenvalues of $A=H_{S}\otimes I_{B1}\otimes I_{B2}$
are plotted. The optimal protocol (black arrows) is obtained by applying
the permutations that lead to monotonically decreasing distribution
of A. (b) Optimal protocol for an X machine (see text). Here the task
is to maximally deplete level no. 1 of the system. For this, we set
$A=\protect\ketbra{1_{S}}{1_{S}}\otimes I_{B1}\otimes I_{B2}$ and
redo the plot. Note that permutation inside each eigenvalue block
(dashed rectangular) has no impact on the final $\left\langle A\right\rangle $.
Thus, a smaller number of permutations (black arrows) can be used
compared to full sorting of the probability distribution.}
\end{figure}

If the operator is local as in the case a refrigerator $A=H_{c}$,
it is important to write it in the Hilbert space of the whole setup
$H_{c}\to H_{c}\otimes I_{rest}$ where $I_{rest}$ is the identity
operator of rest of the setup. As an example, consider the case of
a qutrit with energy spacings $\omega$ that is being cooled by two
spins with energy spacing $\omega$. All the particles start at thermal
equilibrium with inverse temperature $\beta$. The optimal protocol
is obtained by building a bar plot where the $x$ axis contains the
sorted eigenvalues of A, see Fig. \ref{fig: Optimal-protocol}. Local
operators such as $H_{c}$ exhibit many degeneracies, but their ordering
with respect to each other makes no difference in finding the minimal
value of $\left\langle A\right\rangle $. The $y$ axis in Fig. \ref{fig: Optimal-protocol}
is the probability of populating each eigenstate of $H_{c}\otimes I_{rest}$
according to the initial distribution determined by $\rho_{0}^{tot}$.
If the distribution is monotonically decreasing it implies that $\rho_{0}^{tot}$
and $H_{c}\otimes I_{rest}$ are passive with respect to each other
and $\left\langle A\right\rangle $ is already in its minimal passive
value. However, if the distribution is non monotonically decreasing
as in Fig. \ref{fig: Optimal-protocol}a, it is clear that the needed
unitary is the one that rearranges the distribution into monotonically
decreasing form.

As a second example, we consider an exotic heat machine whose goal
is to deplete the population of the middle level ('1') of the qutrit.
We use the setup shown in Fig. \ref{fig: Optimal-protocol}. This
time the operator of interest is $A=\ketbra{1_{S}}{1_{S}}\otimes I_{B1}\otimes I_{B2}$.
Eigenvalue 1 (0 respectively) stands for all the global states of
the setup in which the middle level of the system is (is not) populated.
As before, we plot the distribution of $A$ as shown in (\ref{fig: Optimal-protocol}b),
and apply sorting permutations to minimize the expectation value of
A

Although a complete sorting of the distribution always provides the
optimal protocol, it may contain many operations that do not affect
the observable of interest $A$. Any operation that between degenerate
states of $A$ has no impact on $\left\langle A\right\rangle $ (i.e.
permutation between states in the same dashed box in Fig. \ref{fig: Optimal-protocol}b).
Thus in some case, as in this example, a partial sorting can lead
to the same optimal performance (same change in $\Delta\left\langle A\right\rangle $)
as shown by the black arrows in Fig. \ref{fig: Optimal-protocol}b.
In general, the partial sorting protocol differs from the full sorting
protocol in the final state of the environment (the two qubits) and
its final system-environment correlation.

\section*{Appendix III The Clausius inequality is not tight for small environments}

As described in \citep{MySecondLawReview} and in the references therein,
by assuming that the environment is initially in Gibbs state $\rho_{0}^{env}=e^{-\beta H^{env}}/Z$
the following \textit{equality} holds

\[
\Delta S^{sys}+\beta\Delta\left\langle H^{env}\right\rangle =D(\rho_{f}|\rho_{f}^{sys}\otimes\rho_{f}^{env})+D(\rho_{f}^{env}|\rho_{0}^{env}).
\]
Since the quantum relative entropy satisfies 
\[
D(x,y)>0\:\text{for \ensuremath{x\neq y}},
\]
we get that if $\rho_{f}\neq\rho_{f}^{sys}\otimes\rho_{f}^{env}$
(there is some correlation buildup) or if $\rho_{f}^{env}\neq\rho_{0}^{env}$
(the environment is changed by the interaction with the system) then
\begin{equation}
\Delta S^{sys}+\beta\Delta\left\langle H^{env}\right\rangle >0,\label{eq: CI lager than zero}
\end{equation}
and the CI cannot be saturated. In the microscopic weak-coupling limit
these two relative entropy terms become negligible. However in small
setups, the environment is often driven far away from equilibrium
and non-Markovian dynamics takes place, these terms can be quite large
so that the bound (\ref{eq: CI lager than zero}) is far from being
tight.

\bibliographystyle{plain}
\bibliography{RaamCite1}

\begin{thebibliography}{10}

\bibitem{mahler07b}
{ M. J. Henrich, F. Rempp, G. Mahler}.
\newblock Quantum thermodynamic otto machines: A spin-system approach.
\newblock {\em Eur. Phys. J.}, 151:157, 2007.

\bibitem{AllahverdyanErgotropy}
{A. E. Allahverdyan, R. Balian, and Th. M. Nieuwenhuizen}.
\newblock "maximal work extraction from finite quantum systems".
\newblock {\em Euro. Phys. Lett.}, 67:{565}, 2004.

\bibitem{k272}
{Amikam Levy and Ronnie Kosloff}.
\newblock {Quantum absorption refrigerator}.
\newblock {\em Phys. Rev. Lett.}, 108:070604, 2012.

\bibitem{RobReview2017}
Giuliano Benenti, Giulio Casati, Keiji Saito, and Robert~S. Whitney.
\newblock Fundamental aspects of steady-state conversion of heat to work at the
  nanoscale.
\newblock {\em Physics Reports}, 694(Supplement C):1 -- 124, 2017.
\newblock Fundamental aspects of steady-state conversion of heat to work at the
  nanoscale.

\bibitem{BrandaoPnasRT2ndLaw}
Fernando Brand{\~a}o, Micha{\l} Horodecki, Nelly Ng, Jonathan Oppenheim, and
  Stephanie Wehner.
\newblock The second laws of quantum thermodynamics.
\newblock {\em Proceedings of the National Academy of Sciences},
  112(11):3275--3279, 2015.

\bibitem{palao13}
Luis~A Correa, Jos{\'e}~P Palao, Gerardo Adesso, and Daniel Alonso.
\newblock Performance bound for quantum absorption refrigerators.
\newblock {\em Physical Review E}, 87(4):042131, 2013.

\bibitem{Correa2014EnhancedSciRep}
Luis~A Correa, Jos{\'e}~P Palao, Daniel Alonso, and Gerardo Adesso.
\newblock Quantum-enhanced absorption refrigerators.
\newblock {\em Scientific reports}, 4:3949, 2014.

\bibitem{Esposito2010SecLaw}
Massimiliano Esposito, Katja Lindenberg, and Christian Van~den Broeck.
\newblock Entropy production as correlation between system and reservoir.
\newblock {\em New Journal of Physics}, 12(1):013013, 2010.

\bibitem{Deffner2018ErrorDetectionAnnealers}
Bart{\l}omiej Gardas and Sebastian Deffner.
\newblock Quantum fluctuation theorem for error diagnostics in quantum
  annealers.
\newblock {\em Scientific reports}, 8(1):17191, 2018.

\bibitem{k122}
Eitan Geva and Ronnie Kosloff.
\newblock {The Quantum Heat Engine and Heat Pump: An Irreversible Thermodynamic
  Analysis of The Three-Level Amplifier}.
\newblock {\em J. Chem. Phys.}, 104:7681--7698, 1996.

\bibitem{Landi_Goes2019zero_temp}
Bruno~O Goes, Carlos~E Fiore, and Gabriel~T Landi.
\newblock Quantum features of entropy production in driven-dissipative
  transitions.
\newblock {\em arXiv preprint arXiv:1910.14133}, 2019.

\bibitem{Goold2015review}
John Goold, Marcus Huber, Arnau Riera, L{\'\i}dia del Rio, and Paul Skrzypczyk.
\newblock The role of quantum information in thermodynamics---a topical review.
\newblock {\em Journal of Physics A: Mathematical and Theoretical}, 49:143001,
  2016.

\bibitem{GourRTreview}
Gilad Gour, Markus~P M{\"u}ller, Varun Narasimhachar, Robert~W Spekkens, and
  Nicole~Yunger Halpern.
\newblock The resource theory of informational nonequilibrium in
  thermodynamics.
\newblock {\em Physics Reports}, 583:1--58, 2015.

\bibitem{horodecki2013fundamental}
Micha{\l} Horodecki and Jonathan Oppenheim.
\newblock Fundamental limitations for quantum and nanoscale thermodynamics.
\newblock {\em Nature communications}, 4:2059, 2013.

\bibitem{IBMintroWeb}
https://www.research.ibm.com/ibm q/technology/experience/.

\bibitem{Jarzynski2017_PRX_strong_coupling}
Christopher Jarzynski.
\newblock Stochastic and macroscopic thermodynamics of strongly coupled
  systems.
\newblock {\em Physical Review X}, 7(1):011008, 2017.

\bibitem{Jennings2010ReverseFlow}
David Jennings and Terry Rudolph.
\newblock Entanglement and the thermodynamic arrow of time.
\newblock {\em Physical Review E}, 81(6):061130, 2010.

\bibitem{lenard1978Gibbs}
A~Lenard.
\newblock Thermodynamical proof of the gibbs formula for elementary quantum
  systems.
\newblock {\em Journal of Statistical Physics}, 19:575, 1978.

\bibitem{Lostaglio2019RTreview}
Matteo Lostaglio.
\newblock An introductory review of the resource theory approach to
  thermodynamics.
\newblock {\em Reports on Progress in Physics}, 82(11):114001, 2019.

\bibitem{LostaglioRudolphCohConstraint}
Matteo Lostaglio, David Jennings, and Terry Rudolph.
\newblock Description of quantum coherence in thermodynamic processes requires
  constraints beyond free energy.
\newblock {\em Nature communications}, 6:6383, 2015.

\bibitem{Marshall1979MajorizationBook}
Albert~W Marshall, Ingram Olkin, and Barry~C Arnold.
\newblock {\em Inequalities: theory of majorization and its applications},
  volume 143.
\newblock Springer, 1979.

\bibitem{maslennikov2019quantum}
Gleb Maslennikov, Shiqian Ding, Roland Habl{\"u}tzel, Jaren Gan, Alexandre
  Roulet, Stefan Nimmrichter, Jibo Dai, Valerio Scarani, and Dzmitry
  Matsukevich.
\newblock Quantum absorption refrigerator with trapped ions.
\newblock {\em Nature communications}, 10(1):202, 2019.

\bibitem{merali2017NatureNewsRev}
Zeeya Merali.
\newblock The new thermodynamics: how quantum physics is bending the rules.
\newblock {\em Nature News}, 551(7678):20, 2017.

\bibitem{JanetPRE2017_strong_coupling}
H.~J.~D. Miller and J.~Anders.
\newblock Entropy production and time asymmetry in the presence of strong
  interactions.
\newblock {\em Phys. Rev. E}, 95:062123, Jun 2017.

\bibitem{MarkAbsorptionReview}
Mark~T. Mitchison.
\newblock Quantum thermal absorption machines: refrigerators, engines and
  clocks.
\newblock {\em Contemporary Physics}, 0(0):1--24, 2019.

\bibitem{mitchison2016realising}
Mark~T Mitchison, Marcus Huber, Javier Prior, Mischa~P Woods, and Martin~B
  Plenio.
\newblock Realising a quantum absorption refrigerator with an atom-cavity
  system.
\newblock {\em Quantum Science and Technology}, 1(1):015001, 2016.

\bibitem{MitchisonHuber2015CoherenceAssitedCooling}
Mark~T Mitchison, Mischa~P Woods, Javier Prior, and Marcus Huber.
\newblock Coherence-assisted single-shot cooling by quantum absorption
  refrigerators.
\newblock {\em New Journal of Physics}, 17:115013, 2015.

\bibitem{WolgangSqueezedErgotropy}
Wolfgang Niedenzu, Victor Mukherjee, Arnab Ghosh, Abraham~G Kofman, and Gershon
  Kurizki.
\newblock Universal thermodynamic limit of quantum engine efficiency.
\newblock {\em arXiv preprint arXiv:1703.02911}, 2017.

\bibitem{Wolgang2018passivityCI}
Wolfgang Niedenzu, Victor Mukherjee, Arnab Ghosh, Abraham~G Kofman, and Gershon
  Kurizki.
\newblock Quantum engine efficiency bound beyond the second law of
  thermodynamics.
\newblock {\em Nature communications}, 9(1):165, 2018.

\bibitem{MartiWorkCorr}
Mart\'{\i} Perarnau-Llobet, Karen~V. Hovhannisyan, Marcus Huber, Paul
  Skrzypczyk, Nicolas Brunner, and Antonio Ac\'{\i}n.
\newblock Extractable work from correlations.
\newblock {\em Phys. Rev. X}, 5:041011, Oct 2015.

\bibitem{PeresBook}
Asher Peres.
\newblock {\em Quantum theory: concepts and methods}, volume~57.
\newblock Springer Science \& Business Media, 2006.

\bibitem{EspositoArakiLeib}
Krzysztof Ptaszy\ifmmode~\acute{n}\else \'{n}\fi{}ski and Massimiliano
  Esposito.
\newblock Entropy production in open systems: The predominant role of
  intraenvironment correlations.
\newblock {\em Phys. Rev. Lett.}, 123:200603, Nov 2019.

\bibitem{pusz78}
W.~Pusz and S.L. Wornwicz.
\newblock Passive states and kms states for general quantum systems.
\newblock {\em Commun. Math. Phys.}, 58:273, 1978.

\bibitem{Sagawa2012second}
Takahiro Sagawa.
\newblock Second law-like inequalities with quantum relative entropy: An
  introduction.
\newblock {\em Lectures on Quantum Computing, Thermodynamics and Statistical
  Physics}, 8:127, 2012.

\bibitem{Paternostro2017EntProdWigner}
Jader~P Santos, Gabriel~T Landi, and Mauro Paternostro.
\newblock Wigner entropy production rate.
\newblock {\em Physical review letters}, 118(22):220601, 2017.

\bibitem{Seifert2012StochasticReview}
Udo Seifert.
\newblock Stochastic thermodynamics, fluctuation theorems and molecular
  machines.
\newblock {\em Reports on Progress in Physics}, 75(12):126001, 2012.

\bibitem{SeifertPRL2016_strong_coupling}
Udo Seifert.
\newblock First and second law of thermodynamics at strong coupling.
\newblock {\em Physical review letters}, 116(2):020601, 2016.

\bibitem{sekimotoStochEnergBook}
Ken Sekimoto.
\newblock {\em Stochastic energetics}, volume 799.
\newblock Springer, 2010.

\bibitem{Esposito_PRE2017_strong_coupling}
Philipp Strasberg and Massimiliano Esposito.
\newblock Stochastic thermodynamics in the strong coupling regime: An
  unambiguous approach based on coarse graining.
\newblock {\em Phys. Rev. E}, 95:062101, Jun 2017.

\bibitem{Landi2019landauer_zero_temp}
Andre~M Timpanaro, Jader~P Santos, and Gabriel~T Landi.
\newblock Landauer's principle at zero temperature.
\newblock {\em arXiv preprint arXiv:1911.00910}, 2019.

\bibitem{MySecondLawReview}
Raam Uzdin.
\newblock The second law and beyond in microscopic quantum setups.
\newblock {\em As a chapter of: F. Binder, L. A. Correa, C. Gogolin, J. Anders,
  and G. Adesso (eds.), "Thermodynamics in the quantum regime - Recent Progress
  and Outlook", (Springer International Publishing). arXiv: 1805.02065}.

\bibitem{uzdin2019IBMexp}
Raam Uzdin and Nadav Katz.
\newblock Experimental detection of microscopic environments using
  thermodynamic observables.
\newblock {\em arXiv preprint arXiv:1908.08968}, 2019.

\bibitem{RUswap}
Raam Uzdin and Ronnie Kosloff.
\newblock {The multilevel four-stroke swap engine and its environment}.
\newblock {\em New Journal of Physics}, {16}:{095003}, {2014}.

\bibitem{GlobalPassivity}
Raam Uzdin and Saar Rahav.
\newblock Global passivity in microscopic thermodynamics.
\newblock {\em Phys. Rev. X}, 8:021064, 2018.

\bibitem{SaiJanetReview}
Sai Vinjanampathy and Janet Anders.
\newblock Quantum thermodynamics.
\newblock {\em Contemporary Physics}, 57(4):545--579, 2016.

\end{thebibliography}

\end{document}